\documentclass[showpacs,aps,prd,nofootinbib,showkeys,unsortedaddress,twocolumn,floatfix]{revtex4-1}
\pdfoutput=1

\usepackage{bm}
\usepackage{amsmath}
\usepackage{graphicx}
\usepackage{subfigure}
\usepackage[usenames,dvipsnames]{color}
\definecolor{darkblue}{RGB}{0,0,196}
\usepackage[colorlinks=true,linkcolor=darkblue,citecolor=darkblue,urlcolor=darkblue]{hyperref}

\usepackage{setspace}
\usepackage{footmisc}
\usepackage[makeroom]{cancel}
\usepackage{comment}

\newcommand{\intdP}{\int\!dP}

\def\be{\begin{equation}}
\def\ee{\end{equation}}
\def\ba{\begin{eqnarray}}
\def\ea{\end{eqnarray}}

\begin{document}

\title{Quasiparticle anisotropic hydrodynamics for central collisions}

\author{Mubarak Alqahtani} 
\author{Mohammad Nopoush} 
\author{Michael Strickland} 
\affiliation{Department of Physics, Kent State University, Kent, OH 44242 United States}

\begin{abstract}
We use quasiparticle anisotropic hydrodynamics to study an azimuthally-symmetric boost-invariant quark-gluon plasma including the effects of both shear and bulk viscosities. In quasiparticle anisotropic hydrodynamics, a single finite-temperature quasiparticle mass is introduced and fit to the lattice data in order to implement a realistic equation of state.  We compare results obtained using the quasiparticle method with the standard method of imposing the equation of state in anisotropic hydrodynamics and viscous hydrodynamics.  Using these three methods, we extract the primordial particle spectra, total number of charged particles, and average transverse momentum for various values of the shear viscosity to entropy density ratio $\eta/s$.  We find that the three methods agree well for small shear viscosity to entropy density ratio, $\eta/s$, but differ at large $\eta/s$.  We find, in particular, that when using standard viscous hydrodynamics, the bulk-viscous correction can drive the primordial particle spectra negative at large $p_T$ which is clearly unphysical.  Such a behavior is not seen in either anisotropic hydrodynamics approach, irrespective of the value of $\eta/s$.
\end{abstract}

\date{\today}



\maketitle

\section{Introduction}
\label{sec:intro}

Ultrarelativistic heavy-ion collision
experiments at the Relativistic Heavy Ion Collider (RHIC)
and the Large Hadron Collider (LHC) study the quark-gluon plasma (QGP) using high energy nuclear collisions. The collective behavior seen in these experiments is quite successfully described by relativistic fluid dynamics. In early works, relativistic ideal hydrodynamics was applied assuming the QGP to behave like a perfect fluid \cite{Huovinen:2001cy,Hirano:2002ds,Kolb:2003dz}. Later on, to include the dissipative (viscous) effects, viscous hydrodynamics has been applied \cite{Muronga:2001zk,Muronga:2003ta,Muronga:2004sf,Heinz:2005bw,Baier:2006um,Romatschke:2007mq,Baier:2007ix,Dusling:2007gi,Luzum:2008cw,Song:2008hj,Heinz:2009xj,Bozek:2009ty,Bozek:2009dw,El:2009vj,PeraltaRamos:2009kg,PeraltaRamos:2010je,Denicol:2010tr,Denicol:2010xn,Schenke:2010rr,Schenke:2011tv,Bozek:2011wa,Bozek:2011ua,Niemi:2011ix,Niemi:2012ry,Bozek:2012qs,Denicol:2012cn,Denicol:2012es,PeraltaRamos:2012xk,Calzetta:2014hra,Denicol:2014vaa,Florkowski:2015lra,Ryu:2015vwa,Niemi:2015voa,Niemi:2015bpj}. Recently, due to the large momentum anisotropies generated during heavy-ion collisions, a new framework called anisotropic hydrodynamics has been developed \cite{Martinez:2010sc,Florkowski:2010cf,Ryblewski:2010bs,Martinez:2010sd,Ryblewski:2011aq,Florkowski:2011jg,Martinez:2012tu,Ryblewski:2012rr,Florkowski:2012as,Florkowski:2013uqa,Ryblewski:2013jsa,Bazow:2013ifa,Tinti:2013vba,Florkowski:2014bba,Florkowski:2014txa,Nopoush:2014pfa,Denicol:2014mca,Nopoush:2014pfa,Tinti:2015xra,Bazow:2015cha,Nopoush:2015yga,Bazow:2015zca,Alqahtani:2015qja,Florkowski:2015cba,Molnar:2016vvu} (for a recent review, see Ref.~\cite{Strickland:2014pga}). This new framework has been compared to traditional viscous hydrodynamics in many ways. For boost-invariant and transversely homogeneous systems, by comparing to exact solutions it has been shown that anisotropic hydrodynamics more accurately describes the dynamics in all cases considered \cite{Florkowski:2013lza,Florkowski:2013lya,Bazow:2013ifa,Florkowski:2014sfa,Denicol:2014mca,Tinti:2015xra}. In addition, it has been shown that anisotropic hydrodynamics best reproduces exact solutions of Boltzmann equation subject to 1+1d Gubser flow \cite{Nopoush:2014qba,Denicol:2014xca,Denicol:2014tha}. Finally, we also mention that it has been shown that anisotropic hydrodynamics shows better agreement with data from ultracold Fermi gases experiments than viscous hydrodynamics \cite{Bluhm:2015bzi,Bluhm:2015raa}.

The anisotropic hydrodynamics program is now focused on making phenomenological predictions for heavy-ion physics, including anisotropic freeze-out and a realistic lattice-based equation of state (EoS)~\cite{Nopoush:2015yga}.  In a recent paper it was demonstrated how to impose a realistic EoS assuming approximate conformality of the QGP~\cite{Nopoush:2015yga}.  In Ref.~\cite{Alqahtani:2015qja} a different method for imposing a realistic EoS was proposed in which the non-conformality of the QGP is taken into account by modeling the QGP as a gas of massive quasiparticles with temperature-dependent masses.  This quasiparticle approach is motivated by perturbative results such as hard thermal loop (HTL) resummation, where the quarks and gluons can have temperature-dependent masses \cite{Weldon:1982aq,Weldon:1982bn,Braaten:1989mz,Blaizot:2001nr,Andersen:1999fw,Andersen:2003zk,Andersen:2004fp,Andersen:2011sf,Haque:2014rua}.  In the quasiparticle anisotropic hydrodynamics framework one introduces a single-finite temperature mass which is fit to available lattice data for the QCD EoS.  Once $m(T)$ is determined, the realistic EoS together with the non-equilibrium energy momentum tensor can be used to derive the dynamical equations for such a quasiparticle gas using Boltzmann equation~\cite{Alqahtani:2015qja}.  In this work, we extend the previous 0+1d work of Ref.~\cite{Alqahtani:2015qja} to 1+1d and we use ``anisotropic Cooper-Frye freeze-out'' to compute the primordial particle spectra.

Here we compare results of quasiparticle anisotropic hydrodynamics to the standard anisotropic hydrodynamics \cite{Nopoush:2015yga} and second-order viscous hydrodynamics.  We will refer to the three methods considered herein as ``aHydroQP'', ``aHydro'' and ``vHydro'', respectively. For our comparisons, we derive the dynamical equations necessary for all cases for a 1+1d system which is azimuthally-symmetric and boost invariant.  We then specialize to a relaxation-time approximation for collisional kernel and solve the partial differential equations for each approach numerically.  We then present comparisons of the total number of charged particles $N_{\rm chg}$, the average transverse momentum $\langle p_T \rangle$ for pions, kaons, and protons, and the differential pion, kaon, and proton spectra predicted by each approach.  We find that the three methods agree well for small shear viscosity to entropy density ratio, $\eta/s$, but differ at large $\eta/s$.  We find, in particular, that when using standard viscous hydrodynamics, the bulk-viscous correction can drive the primordial particle spectra negative at large $p_T$.  Such a behavior is not seen in either anisotropic hydrodynamics approach, irrespective of the value of $\eta/s$.

The structure of the paper is as follows. In Sec.~\ref{sec:conventions}, we specify the notation and conventions used in the paper.  In Sec.~\ref{sec:setup}, we review the necessary setup including basis vectors necessary in different cases, the anisotropic distribution function, and the lattice based equation of state.  In Sec.~\ref{sec:boltzman}, the Boltzmann equation and its generalization to quasiparticles with temperature-dependent masses is discussed. In Sec.~\ref{sec:bulk-var}, we derive expressions for the particle four-current, energy density, and components of the pressure by taking different moments of distribution function.  In Sec.~\ref{sec:dynamical-eqs}, the dynamical equations for massive anisotropic hydrodynamics are derived for azimuthally-symmetric boost-invariant systems. In Sec.~\ref{sec:freeze-out} we discuss anisotropic Cooper-Frye freeze-out in the context of leading-order anisotropic hydrodynamics.  In Sec.~\ref{sec:results}, our numerical results obtained using the three methods for central Pb-Pb and p-Pb collisions at LHC energies are presented. Sec.~\ref{sec:conclusions} contains our conclusions and an outlook for the future. In App.~\ref{app:basis} the basis vectors used in the paper are presented. In App.~\ref{app:vhydro} we present details about second-order viscous hydrodynamics equations.  Finally, all necessary identities and function definitions are collected in Apps.~\ref{app:identities} and \ref{app:h-functions}.
 
\section{Conventions and Notation}
\label{sec:conventions}

A parentheses in the indices indicates a symmetrized form, e.g. $A^{(\mu\nu)} \equiv (A^{\mu\nu} + A^{\nu\mu})/2$.  The metric is taken to be in the ``east coast convention'' such that in Minkowski space with $x^\mu\equiv(t,x,y,z)$ the measure is $ds^2=g_{\mu\nu} dx^\mu dx^\nu=dt^2-dx^2-dy^2-dz^2$.  We also use the standard transverse projector, $\Delta^{\mu\nu} \equiv g^{\mu\nu} - u^\mu u^\nu$.  When studying relativistic heavy-ion collisions, it is convenient to transform to Milne coordinates defined by $\tau =\sqrt{t^2 - z^2}$, which is the longitudinal proper time, and $\varsigma ={\rm tanh}^{-1}(z/t)$, which is the longitudinal spacetime rapidity.   For a system which is azimuthally symmetric with respect to the beam-line, it is convenient to transform to polar coordinates in the transverse plane with $r=\sqrt{x^{2}{+}y^{2}}$ and $\phi ={\rm tan}^{-1}(y/x)$.   In this case, the new set of coordinates $x^\mu=(\tau,r,\phi,\varsigma)$ defines polar Milne coordinates.  Finally, the invariant phase space integration measure is defined as
\be
dP \equiv N_{\rm dof} \frac{d^3p}{(2\pi)^3} \frac{1}{E} = \tilde{N} \frac{d^3p}{E} \, ,
\ee
where $\tilde{N} \equiv N_{\rm dof}/(2\pi)^3$ with $N_{\rm dof}$ being the number of degrees of freedom.

\section{Setup}
\label{sec:setup}

In this paper, we derive non-conformal anisotropic hydrodynamics equations for a system of quasiparticles with a temperature-dependent mass. To accomplish this goal, we take moments of Boltzmann equation, appropriate for the system of quasiparticles with thermal mass, to obtain the necessary anisotropic hydrodynamics equations. Using a general set of basis vectors and some simplifying assumptions relevant for 1+1d system, equations are expanded and simplified to the form appropriate for describing a boost-invariant and azimuthally-symmetric QGP. The obtained 1+1d equations are then solved numerically for our tests.

\subsection{Ellipsoidal form}
\label{subsec:distribution-func}

In non-conformal anisotropic hydrodynamics one introduces an anisotropy tensor of the form \cite{Nopoush:2014pfa,Martinez:2012tu}
\be
\Xi^{\mu\nu} = u^\mu u^\nu + \xi^{\mu\nu} - \Delta^{\mu\nu} \Phi \, ,
\label{eq:aniso-tensor1}
\ee
where $u^\mu$ is four-velocity, $\xi^{\mu\nu}$ is a symmetric and traceless tensor, and $\Phi$ is associated with the bulk degree of freedom.  The quantities $u^\mu$, $\xi^{\mu\nu}$, and $\Phi$ are understood to be functions of spacetime and obey $u^\mu u_\mu = 1$, ${\xi^{\mu}}_\mu = 0$, ${\Delta^\mu}_\mu = 3$, and $u_\mu \xi^{\mu\nu} = 0$; therefore, one has ${\Xi^\mu}_\mu = 1 - 3 \Phi$.
At leading order in the anisotropic hydrodynamics expansion one assumes that the one-particle distribution function is of the form
\be
f(x,p) = f_{\rm iso}\!\left(\frac{1}{\lambda}\sqrt{p_\mu \Xi^{\mu\nu} p_\nu}\right) ,
\label{eq:genf}
\ee
where $\lambda$ has dimensions of energy and can be identified with the temperature only in the isotropic equilibrium limit ($\xi^{\mu\nu} = 0$ and $\Phi=0$).\footnote{Herein we assume that the chemical potential is zero.} 
We note that, in practice, $f_{\rm iso}$ need not be a thermal equilibrium distribution.  However, unless one expects there to be a non-thermal distribution at late times, it is appropriate to take $f_{\rm iso}$ to be a thermal equilibrium distribution function of the form $f_{\rm iso}(x) = f_{\rm eq}(x) = (e^x + a )^{-1}$,  where $a= \pm 1$ gives Fermi-Dirac or Bose-Einstein statistics, respectively, and $a=0$ gives Boltzmann statistics. From here on, we assume that the distribution is of Boltzmann form, i.e. $a=0$. 

Since the most important viscous corrections are the diagonal components of the energy-momentum tensor, to good approximation one can assume that \mbox{$\xi^{\mu\nu} = {\rm diag}(0,{\boldsymbol \xi})$} with ${\boldsymbol \xi} \equiv (\xi_x,\xi_y,\xi_z)$ and $\xi^i_i=0$.\footnote{For a 1+1d system this is exact since one only needs the $rr$, $\phi\phi$, and $\varsigma\varsigma$ components of the energy-momentum tensor.}  In this case, expanding the argument of the square root appearing on the right-hand side of Eq.~(\ref{eq:genf}) in the LRF gives
\be
f(x,p) =  f_{\rm eq}\!\left(\frac{1}{\lambda}\sqrt{\sum_i \frac{p_i^2}{\alpha_i^2} + m^2}\right) ,
\label{eq:fform}
\ee
where $i\in \{x,y,z\}$ and the scale parameters $\alpha_i$ are
\be
\alpha_i \equiv (1 + \xi_i + \Phi)^{-1/2} \, .
\label{eq:alphadef}
\ee  
Note that, for compactness, one can collect the three anisotropy parameters into vector $\boldsymbol\alpha \equiv (\alpha_x,\alpha_y,\alpha_z)$. In the isotropic equilibrium limit, where $\xi_i = \Phi = 0$ and $\alpha_i =1$, one has $p_\mu \Xi^{\mu\nu} p_\nu = (p \cdot u)^2 = E^2$  and $\lambda\rightarrow T$ and, hence,
\be
f(x,p)=f_{\rm eq}\!\left(\frac{E}{T(x)}\right) .
\label{eq:feqform}
\ee
Out of the four anisotropy and bulk parameters there are only three independent ones. In practice, we use the three variables $\alpha_i$ as the dynamical anisotropy parameters since, by using Eq.~(\ref{eq:alphadef}) and the tracelessness of $\xi^{\mu\nu}$, one can write $\Phi$ in terms of the anisotropy parameters, $\Phi = \frac{1}{3} \sum_i \alpha_i^{-2} - 1$.

\subsection{Equation of state}
\label{subsec:eos}
Herein we consider a system at finite temperature and zero chemical potential. At asymptotically high temperatures, the pressure of a gas of quarks and gluons approaches the Stefan-Boltzmann (SB) limit.  At the temperatures probed in heavy-ion collisions there are important corrections to the SB limit and at low temperatures the relevant degrees of freedom change from quarks and gluons to hadrons.  The standard way to determine the QGP EoS is to use non-perturbative lattice calculations.  For this purpose, we use an analytic parameterization of lattice data for the QCD interaction measure (trace anomaly), $I_{\rm eq} = {\cal E}_{\rm eq} - 3 {\cal P}_{\rm eq}$, taken from the Wuppertal-Budapest collaboration \cite{Borsanyi:2010cj}. We refer the reader to the reference  \cite{Alqahtani:2015qja} for more details.
 
\subsubsection*{Method 1:  Standard equation of state}

In the standard approach for imposing a realistic EoS in anisotropic hydrodynamics, one derives the necessary equations in the conformal limit and exploits the conformal multiplicative factorization of the components of the energy-momentum tensor~\cite{Martinez:2010sc,Florkowski:2010cf}.  With this method, one relies on the assumption of factorization even in the non-conformal (massive) case.  Such an approach is justified by the smallness of the corrections to factorization in the massive case in the near-equilibrium limit~\cite{Nopoush:2015yga}.  For details concerning this method, we refer the reader to Refs.~\cite{Ryblewski:2012rr,Nopoush:2015yga}.  Although this method is relatively straightforward to implement, it is only approximate since for non-conformal systems there is no longer exact multiplicative factorization of the components of the energy-momentum tensor.  This introduces a theoretical uncertainty which is difficult to quantitatively estimate.

\subsubsection*{Method 2:  Quasiparticle equation of state}

Since the standard method is only approximate, one would like to find an alternative method for imposing a realistic equation of state in an anisotropic system that can be applied for non-conformal systems.  In order to accomplish this goal, we  implement the realistic EoS detailed above by assuming that the QGP can be described as an ensemble of massive quasiparticles with temperature-dependent masses.  As is well-known from the literature \cite{gorenstein1995gluon}, one cannot simply substitute temperature-dependent masses into the thermodynamic functions obtained with constant masses because this would violate thermodynamic consistency.  For an equilibrium system, one can ensure thermodynamic consistency by adding a background contribution to the energy-momentum tensor, i.e.
\be
T^{\mu\nu}_{\rm eq} = T^{\mu\nu}_{\rm kinetic,eq} + g^{\mu\nu} B_{\rm eq}  \, ,
\ee
with $B_{\rm eq}\equiv B_{\rm eq}(T)$ being the additional background contribution.  The kinetic contribution to the energy momentum tensor is given by
\be
T^{\mu\nu}_{\rm kinetic, eq} = \intdP \, p^\mu p^\nu f_{\rm eq}(x,p) \, .
\ee

For an equilibrium Boltzmann gas, the number and entropy densities are unchanged, while, due to the additional background contribution, the energy density and pressure are shifted by $+B_{\rm eq}$ and $-B_{\rm eq}$, respectively, giving
\ba
n_{\rm eq} &=& 4 \pi \tilde{N} T^3 \, \hat{m}_{\rm eq}^2 K_2\left( \hat{m}_{\rm eq}\right) , \label{eq:neq} \nonumber \\
{\cal S}_{\rm eq} &=&4 \pi \tilde{N} T^3 \, \hat{m}_{\rm eq}^2 \Big[4K_2\left( \hat{m}_{\rm eq}\right)+\hat{m}_{\rm eq}K_1\left( \hat{m}_{\rm eq}\right)\Big] ,
\label{eq:Seq} \nonumber \\
{\cal E}_{\rm eq} &=& 4 \pi \tilde{N} T^4 \, \hat{m}_{\rm eq}^2
 \Big[ 3 K_{2}\left( \hat{m}_{\rm eq} \right) + \hat{m}_{\rm eq} K_{1} \left( \hat{m}_{\rm eq} \right) \Big]+B_{\rm eq} \, , 
\label{eq:Eeq} \nonumber \\
 {\cal P}_{\rm eq} &=& 4 \pi \tilde{N} T^4 \, \hat{m}_{\rm eq}^2 K_2\left( \hat{m}_{\rm eq}\right)-B_{\rm eq} \, ,
\label{eq:Peq}
\ea
where $\hat{m}_{\rm eq} = m/T$ with $m$ implicitly depending on the temperature from here on.  In order to fix $B_{\rm eq}$, one can require, for example, the thermodynamic identity
\be
T {\cal S}_{\rm eq} = {\cal E}_{\rm eq} + {\cal P}_{\rm eq} = T \frac{\partial P_{\rm eq}}{\partial T} \, ,
\label{eq:thermoid}
\ee
be satisfied.  Using Eqs.~(\ref{eq:Peq}) and (\ref{eq:thermoid}) one obtains 
\ba
\frac{dB_{\rm eq}}{dT} &=& - \frac{1}{2} \frac{dm^2}{dT} \intdP \, f_{\rm eq}(x,p) \nonumber \\  
&=& -4\pi \tilde{N}m^2 T K_1(\hat{m}_{\rm eq}) \frac{dm}{dT} \, .
\label{eq:BM-matching-eq-1}
\ea
If the temperature dependence of $m$ is known, then Eq.~(\ref{eq:BM-matching-eq-1}) can be used to determine $B_{\rm eq}$.
In practice, in order to determine $m$, one can use the thermodynamic identity
\be
{\cal E}_{\rm eq}+{\cal P}_{\rm eq}=T{\cal S}_{\rm eq} = 4 \pi \tilde{N} T^4 \, \hat{m}_{\rm eq}^3 K_3\left( \hat{m}_{\rm eq}\right) .
\label{eq:meq}
\ee
Using the lattice data parameterization to compute the equilibrium energy density and pressure, one can numerically solve for $m(T)$.  We refer the reader to Ref.~\cite{Alqahtani:2015qja} for more details.

\section{Boltzmann equation and its moments}
\label{sec:boltzman}

In this paper, we derive the necessary hydrodynamical equations by taking the moments of Boltzmann equation.   In this section, we introduce the Boltzmann equation and its different moments for the general case of quasiparticles with a temperature-dependent mass. Then we simplify them for the case of massless particles, when necessary, since the massless equations are used in the standard approach. 

\subsection{Boltzmann Equation}
\label{subsec:thermal-boltzmann}

Generally, the Boltzmann equation for on-shell quasiparticles with temperature dependent mass can be written as ~\cite{Jeon:1995zm,Romatschke:2011qp,Alqahtani:2015qja}

\be
p^\mu \partial_\mu f+\frac{1}{2}\partial_i m^2\partial^i_{(p)} f=-\mathcal{C}[f]\,,
\label{eq:boltz2}
\ee
where $p^\mu\equiv(\sqrt{{\bf p}^2+m^2},\bf p)$ is the on-shell momentum four-vector, $i$ indexes the spatial coordinates, and
$\partial^i_{(p)}\equiv -\partial/\partial p^i$. In the constant mass limit, the above Boltzmann equation simplifies to 
\be
p^\mu \partial_\mu f = -{\cal C}[f]\, .
\label{eq:boltzmanneq}
\ee
The function ${\cal C}[f]$ appearing above is the collisional kernel containing all interactions involved in the dynamics. In what follows, we specialize to the case that the collisional kernel is given by the relaxation-time approximation (RTA), however, the general methods presented here can be applied to any collisional kernel. In RTA, one has
\be
{\cal C}[f]=\frac{p^\mu u_\mu}{\tau_{\rm eq}}(f-f_{\rm eq})\,.
\label{eq:RTA}
\ee
In this relation, $f_{\rm eq}$ denotes the equilibrium one-particle distribution function (\ref{eq:feqform}) and $\tau_{\rm eq}$ is the relaxation time which can depend on spacetime but which we assume to be momentum-independent.  To obtain a realistic model for $\tau_{\rm eq}$, which is valid for massive systems, one can relate $\tau_{\rm eq}$ to the shear viscosity to entropy density ratio as \cite{anderson1974relativistic,Czyz:1986mr}
\be
\eta(T)=\frac{\tau_{\rm eq}(T) {\cal P}_{\rm eq}(T)}{15}\kappa(\hat{m}_{\rm eq})\,.
\ee
In this formula the function $\kappa(x)$ is defined as
\ba 
\kappa(x) &\equiv& x^3 \bigg[\frac{3}{x^2}\frac{K_3(x)}{K_2(x)}-\frac{1}{x}+\frac{K_1(x)}{K_2(x)}
\nonumber \\ && \;\;\;\;
-\frac{\pi}{2}\frac{1-xK_0(x)L_{-1}(x)-xK_1(x)L_0(x)}{K_2(x)}\bigg] , \hspace{5mm}
\ea
where $K_n(x)$ are modified Bessel functions of second kind and $L_n(x)$ are modified Struve functions.  Assuming that the ratio of the shear viscosity to entropy density, $\eta/{\cal S}_{\rm eq}\equiv \bar{\eta}$, is held fixed during the evolution and using the thermodynamic relation ${\cal E}_{\rm eq}+{\cal P}_{\rm eq}=T{\cal S}_{\rm eq}$ one obtains
\be 
\tau_{\rm eq}(T)=\frac{15 \bar{\eta}}{\kappa(\hat{m}_{\rm eq})T}\bigg(1+\frac{{\cal E}_{\rm eq}(T)}{{\cal P}_{\rm eq}(T)}\bigg) .
\label{eq:teq}
\ee
Note that, $\lim_{m\rightarrow 0}  \kappa(\hat{m}_{\rm eq})\rightarrow12$, giving 
\be
\lim_{m\rightarrow 0} \tau_{\rm eq}(T)= \frac{5 \eta}{4{\cal P}_{\rm eq}(T)} \, .
\label{eq:teq0}
\ee
%
\subsection{Moments of Boltzmann Equation}
\label{subsec:boltzmann-moments}

By calculating moments of Boltzmann equation one obtains evolution equations for tensors of different ranks, with the zeroth moment giving the evolution of particle four-current, the first moment giving the evolution of the energy-momentum tensor, and the second-moment describing the evolution of a particular rank three tensor.
Taking the zeroth, first, and second moments of Boltzmann equation gives, respectively 
\ba
\partial_\mu J^\mu&=&-\intdP \, {\cal C}[f]\, , \label{eq:J-conservation} \\
\partial_\mu T^{\mu\nu}&=&-\intdP \, p^\nu {\cal C}[f]\, , \label{eq:T-conservation} \\
\partial_\mu {\cal I}^{\mu\nu\lambda}- J^{(\nu} \partial^{\lambda)} m^2 &=&-\intdP \, p^\nu p^\lambda{\cal C}[f]\, \label{eq:I-conservation},   
\ea 
where the particle four-current $J^\mu$, energy-momentum tensor $T^{\mu\nu}$, and the rank-three tensor ${\cal I}^{\mu\nu\lambda}$ are given by
\ba
J^\mu &\equiv& \intdP \, p^\mu f(x,p)\, , \label{eq:J-int} \\
T^{\mu\nu}&\equiv& \intdP \, p^\mu p^\nu f(x,p)+B g^{\mu\nu}, \label{eq:T-int}\\
{\cal I}^{\mu\nu\lambda} &\equiv& \intdP \, p^\mu p^\nu p^\lambda  f(x,p) \, .
\label{eq:I-int}
\ea
We note that we have introduced the non-equilibrium background field $B\equiv B({\boldsymbol\alpha},\lambda)$, which is the analogue of the equilibrium background $B_{\rm eq}$ in order to guarantee that the correct equilibrium limit of $T^{\mu\nu}$ is obtained.  In the process of the derivation one finds that, in order to write the energy momentum conservation in the form given in Eq.~(\ref{eq:T-conservation}), there must be a differential equation relating $B$ and the thermal mass \cite{Alqahtani:2015qja}
\be
\partial_\mu B = -\frac{1}{2} \partial_\mu m^2 \intdP  f(x,p)\,.
\label{eq:BM-matching}
\ee
In practice, one can use (\ref{eq:BM-matching}) to write the derivative of $B$ with respect to any variable in terms of the derivative of the thermal mass times the $E^{-1}$ moment of the non-equilibrium distribution function.

\section{Bulk variables}
\label{sec:bulk-var}

In this section, the bulk variables necessary (number density, energy density, and the pressures) are calculated by taking projections of $J^\mu$ and $T^{\mu\nu}$.

\subsection{Particle four-current}
\label{subsec:4current}
The particle four-current $J^\mu\equiv(n,\bf{J})$ is defined in Eq.~(\ref{eq:J-int}). Using Eqs.~(\ref{eq:fform}) and (\ref{eq:J-int}) one has
\be
J^\mu=(n,{\bf 0})=nu^\mu \, ,
\label{eq:J-exp2}
\ee
where $n=\alpha n_{\rm eq}(\lambda,m)$ and $\alpha\equiv\alpha_x \alpha_y \alpha_z$.

\subsection{Energy-momentum tensor}  
\label{subsec:T-tensor}

The energy-momentum tensor $T^{\mu\nu}$ is defined in Eq.~(\ref{eq:T-int}). Expanding it using the basis vectors one obtains
\be
T^{\mu\nu}={\cal E}u^\mu u^\nu+{\cal P}_x X^\mu X^\nu+{\cal P}_y Y^\mu Y^\nu+{\cal P}_z Z^\mu Z^\nu \, .
\label{eq:T-expan}
\ee

\subsubsection{Quasiparticle method}

Using Eqs.~(\ref{eq:fform}), (\ref{eq:T-int}), and (\ref{eq:T-expan}) and taking projections of $T^{\mu\nu}$ one can obtain the energy density and the components of pressure
\ba
{\cal E} &=& {\cal H}_3({\boldsymbol\alpha},\hat{m}) \, \lambda^4+B \, ,\nonumber \\
{\cal P}_x &=& {\cal H}_{3x}({\boldsymbol\alpha},\hat{m}) \, \lambda^4-B \, ,\nonumber \\
{\cal P}_y &=& {\cal H}_{3y}({\boldsymbol\alpha},\hat{m}) \, \lambda^4-B \, ,\nonumber \\
{\cal P}_z &=& {\cal H}_{3L}({\boldsymbol\alpha},\hat{m}) \, \lambda^4-B \, ,
\label{eq:bulk_var}
\ea
where $\hat{m} \equiv m/\lambda$.

\subsubsection{Standard method}

For a massless conformal Boltzmann gas, one has ${\cal E}_{\rm eq}(\lambda) = 24 \pi \tilde{N}\lambda^4$ and ${\cal P}_{\rm eq}(\lambda) = 8\pi \tilde{N}\lambda^4$.  Using these relations, one can rewrite Eqs.~(\ref{eq:bulk_var}) in the standard case, by taking the massless limit, $m \rightarrow0$, and hence $B\rightarrow 0$ 
\ba
{\cal E} &=& {\cal E}_{\rm eq}(\lambda) \, \hat{{\cal H}}_3({\boldsymbol\alpha}) \,  \, ,\nonumber \\
{\cal P}_x &=& {\cal P}_{\rm eq}(\lambda) \,\hat{{\cal H}}_{3x}({\boldsymbol\alpha}) \,  \, ,\nonumber \\
{\cal P}_y &=& {\cal P}_{\rm eq}(\lambda) \,\hat{{\cal H}}_{3y}({\boldsymbol\alpha}) \,  \, ,\nonumber \\
{\cal P}_z &=& {\cal P}_{\rm eq}(\lambda) \,\hat{{\cal H}}_{3L}({\boldsymbol\alpha}) \,  .
\label{eq:bulk_var_massless_fac}
\ea
These formulas suggest that, in order to impose a realistic EoS, one only has to replace ${\cal E}_{\rm eq}(\lambda)$ and ${\cal P}_{\rm eq}(\lambda)$ by the results obtained from lattice QCD calculations.

\section{Dynamical equations}
\label{sec:dynamical-eqs}

In order to obtain the dynamical equations from Eqs.~(\ref{eq:J-conservation})-(\ref{eq:I-conservation}), one needs to impose the RTA collisional kernel. Enforcing energy-momentum tensor conservation leads to an extra matching (constraint) equation. In total, one has the following four general dynamical equations
\ba
\partial_\mu J^\mu &=&\frac{1}{\tau_{\rm eq}}(n_{\rm eq}-n) \, ,  \\
\partial_\mu T^{\mu\nu}&=&0\, ,  \\
\partial_\mu {\cal I}^{\mu\nu\lambda}-J^{(\nu} \partial^{\lambda)} m^2 &=& \frac{u_\mu}{\tau_{\rm eq}}( {\cal I}^{\mu\nu\lambda}_{\rm eq}- {\cal I}^{\mu\nu\lambda})\,,  \label{eq:2ndmom} \\
{\cal E}_{\rm kinetic}&=&{\cal E}_{\rm kinetic,eq}\,.
\ea
Using the tensor decomposition of $J_\mu$, $T_{\mu\nu}$, and ${\cal I}_{\mu\nu\lambda}$ in the basis vectors appropriate for a boost-invariant and azimuthally symmetric 1+1d system, one can expand the expressions above to obtain the final form of the equations \cite{Nopoush:2014pfa,Alqahtani:2015qja}. Choosing two equations from the first moment, three from the second moment, together with the matching condition we end up with six equations for six independent variables $\boldsymbol\alpha$, $\lambda$, $T$, and $\theta_\perp$. The non-trivial equations from the first moment are
\ba 
&& D_u{\cal E}+{\cal E}\theta_u+ {\cal P}_x D_x \theta_{\perp}
\nonumber \\ 
&& \hspace{1cm} + \frac{1}{r}{\cal P}_y \sinh\theta_{\perp} +\frac{1}{\tau}{\cal P}_z\cosh\theta_{\perp} = 0 \, , \label{eq:1st-mom-1}\\
&& D_x {\cal P}_x+{\cal P}_x\theta_x +{\cal E} D_u\theta_{\perp} \nonumber \\ 
&&  \hspace{1cm} -\frac{1}{r}{\cal P}_y\cosh\theta_{\perp}-\frac{1}{\tau}{\cal P}_z\sinh\theta_{\perp} = 0
\label{eq:1st-mom-2} \,.
\ea
The convective derivatives $D_\alpha$ and divergences $\theta_\alpha$, with $\alpha\in\{u,x,y,z\}$, are defined in App.~\ref{app:identities}. Depending on the model, one can replace the bulk variables from (\ref{eq:bulk_var}) or (\ref{eq:bulk_var_massless_fac}) in the above equations to obtain the final form of dynamical equations for massive quasiparticle or massless standard models, respectively.
 
Also, taking the $XX$-, $YY$-, and $ZZ$- projections of the second moment equation (\ref{eq:2ndmom}), one obtains
\ba 
\frac{D_u {\cal I}_x}{{\cal I}_x} + \theta_u + 2 D_x \theta_{\perp}
&=& \frac{1}{\tau_{\rm eq}} ( \frac{{\cal I}_{\rm eq}}{{\cal I}_x} - 1 ) \, ,  \\
\frac{D_u {\cal I}_y}{{\cal I}_y} + \theta_u + \frac{2}{r} \sinh \theta_{\perp}
&=& \frac{1}{\tau_{\rm eq}} ( \frac{{\cal I}_{\rm eq}}{{\cal I}_y} - 1 ) \, , \\
\frac{D_u {\cal I}_z}{{\cal I}_z} + \theta_u + \frac{2}{\tau} \cosh \theta_{\perp}
&=& \frac{1}{\tau_{\rm eq}} ( \frac{{\cal I}_{\rm eq}}{{\cal I}_z} - 1 ) \, . 
\ea
Evaluating the necessary integrals using the distribution function (\ref{eq:fform}), one finds
\ba
{\cal I}_u &=& \Big(\sum_i \alpha_i^2\Big) \alpha \, {\cal I}_{\rm eq}(\lambda,m) + \alpha m^2 n_{\rm eq}(\lambda,m) \, ,
\\
{\cal I}_i &=& \alpha \, \alpha_i^2 \, {\cal I}_{\rm eq}(\lambda,m) \, ,
\label{eq:I-i}
\ea
with ${\cal I}_{\rm eq}(\lambda,m) =  4 \pi {\tilde N} \lambda^5 \hat{m}^3 K_3(\hat{m})$. In the massless case, they simplify to 
\ba
{\cal I}_u &=& \Big(\sum_i \alpha_i^2\Big) \alpha \, {\cal I}_{\rm eq}(\lambda)  \, ,
\\
{\cal I}_i &=& \alpha \, \alpha_i^2 \, {\cal I}_{\rm eq}(\lambda) \, ,
\ea
with ${\cal I}_{\rm eq}(\lambda) =  32 \pi {\tilde N} \lambda^5$. Finally, the matching condition is
\ba
{\cal H}_3 ({\boldsymbol\alpha},\hat{m})\lambda^4 &=& {\cal H}_{3,\rm eq}(\hat{m}_{\rm eq}) T^4,
\ea
where
\be
{\cal H}_{3,\rm eq}(\hat{m}_{\rm eq})\equiv \lim_{\substack{\lambda\rightarrow T \\ \boldsymbol\alpha\rightarrow 1 }}  {\cal H}_3 ({\boldsymbol\alpha},\hat{m}). 
\ee
%

\section{Anisotropic Freeze-out}
\label{sec:freeze-out}

We now turn to the topic of hadronic freeze-out.  Our technique will be to perform ``anisotropic Cooper-Frye freeze-out'' using Eq.~(\ref{eq:genf}) as the form for the one-particle distribution function. This is different than the typical freeze-out prescription used in viscous hydrodynamics in which one takes into account the dissipative correction to the equilibrium distribution function only at linear order in a Taylor expansion around equilibrium. One immediate benefit of performing anisotropic freeze-out using Eq.~(\ref{eq:genf}) is that, with this form, one is guaranteed that the one-particle distribution function is positive-definite in all regions in phase space.

In practice, we start from the standard freeze-out integral
\be
N=\int_{\Sigma} d^3\Sigma_\mu J^\mu  \, ,
\label{eq:particle-number1} 
\ee
where $\Sigma$ is the three-dimensional freeze-out hypersurface defining the boundary of the four-dimensional volume occupied by the fluid, $d^3\Sigma^\mu$ is the surface normal vector, and $J^\mu$ is the particle four-current.  Due to the presence of momentum-space anisotropies, one cannot simply use the momentum scale $\lambda$ when defining the freeze-out hypersurface $\Sigma$.  Instead, one should use the energy density, from which one can obtain the effective freeze-out temperature $T_{\rm FO} \equiv T_{\rm eff,FO}= T({\cal E}_{\rm FO})$ where $T({\cal E})$ is obtained using our realistic EoS. After identifying $\Sigma$, we follow the parametrization presented in \cite{Nopoush:2015yga}. 

Unlike \cite{Nopoush:2015yga}, here we take into account the breaking of conformality.  The only change required is in the distribution function itself.
Parameterizing the particle momentum in the lab frame as 
\ba
p^\mu\equiv(m_\perp \cosh y,p_\perp \cos\varphi,p_\perp \sin\varphi,m_\perp \sinh y)\, ,
\label{eq:ptl-mom}
\ea
where $m_\perp = \sqrt{p_\perp^2+m^2}$, $y = \tanh^{-1}(p^z/p^0)$ is the particle's rapidity, and $\varphi$ is the particle's azimuthal angle.
In order to set up the distribution function, having $p^\mu$ defined in Eq.~(\ref{eq:ptl-mom}), one can use Eqs.~(\ref{eq:aniso-tensor1}) and (\ref{eq:genf}) to find in both the quasiparticle and standard cases
\begin{widetext}
\ba
p^\mu \Xi_{\mu\nu} p^\nu &=& 
(1+\Phi)\Big[m_\perp \cosh \theta _\perp \cosh
   (y-\varsigma)-p_\perp \sinh \theta _\perp \cos (\phi -\varphi )\Big]^2
   \nonumber \\ 
   &+& \xi _x\,\Big[m_\perp \sinh \theta _\perp \cosh (y-\varsigma)-p_\perp \cosh \theta
   _\perp \cos (\phi -\varphi )\Big]^2
   \nonumber \\
   &+& \xi_z\, m_\perp^2  \sinh^2(y-\varsigma)
   +\xi_y\,\, p_\perp^2  \sin^2(\phi -\varphi ) 
    - \Phi m^2 \,.
\label{eq:pxipqu}
\ea
\end{widetext}
Note that  $ p^\mu \Xi_{\mu\nu} p^\nu $ is Lorentz invariant, and by going to LRF, one can show that this is positive definite in any frame.

\begin{figure}[t]
\includegraphics[width=0.9\linewidth]{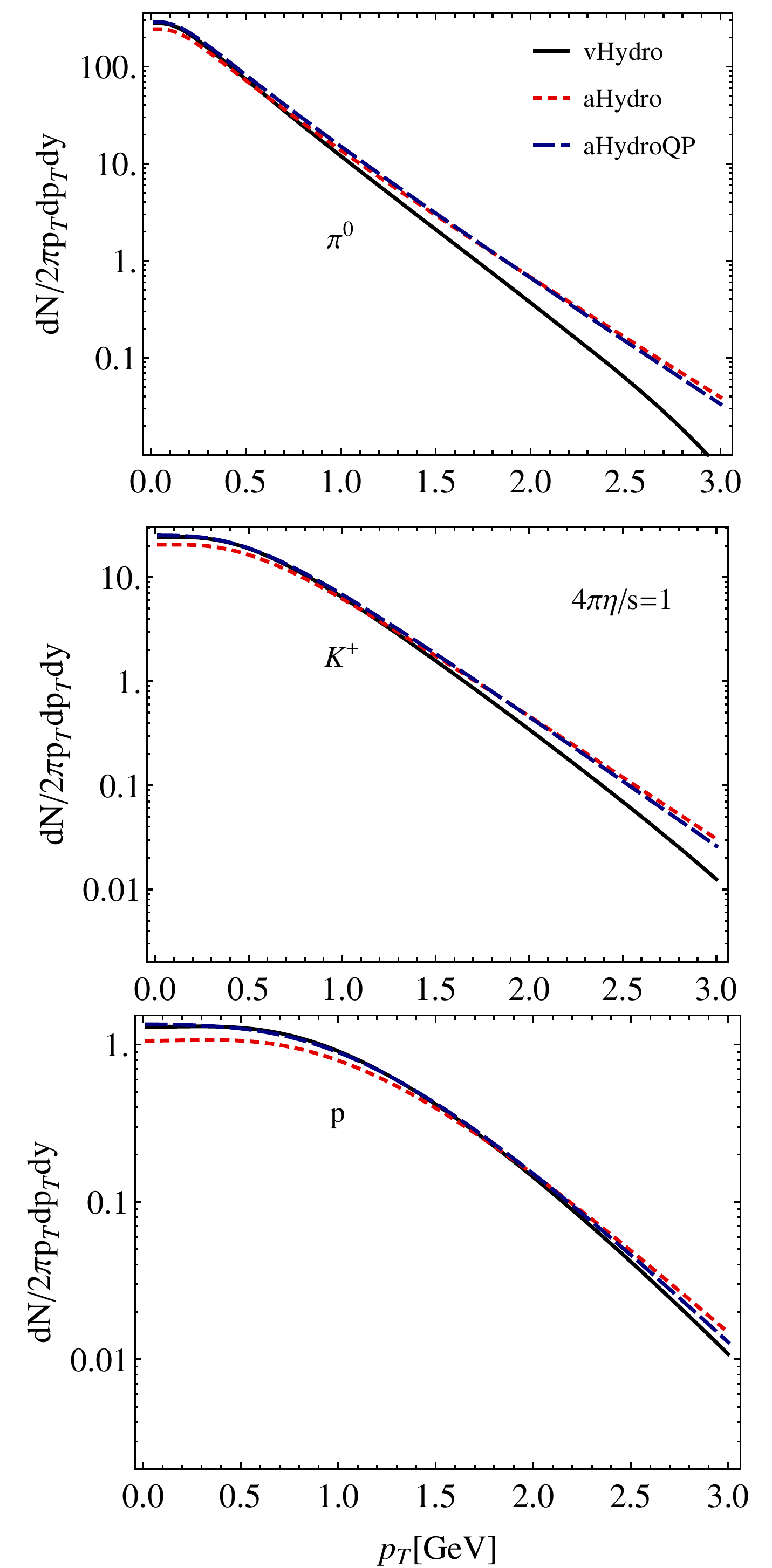}
\caption{Comparison of the neutral pions, kaons ($K^+$), and protons spectra as a function of transverse momentum $ p_T$ obtained using aHydroQP, aHydro and  vHydro. The shear viscosity to entropy density ratio is $ 4 \pi \eta / s =1$.}
\label{fig:spectrum1}
\end{figure}

\begin{figure}[t]
\includegraphics[width=0.9\linewidth]{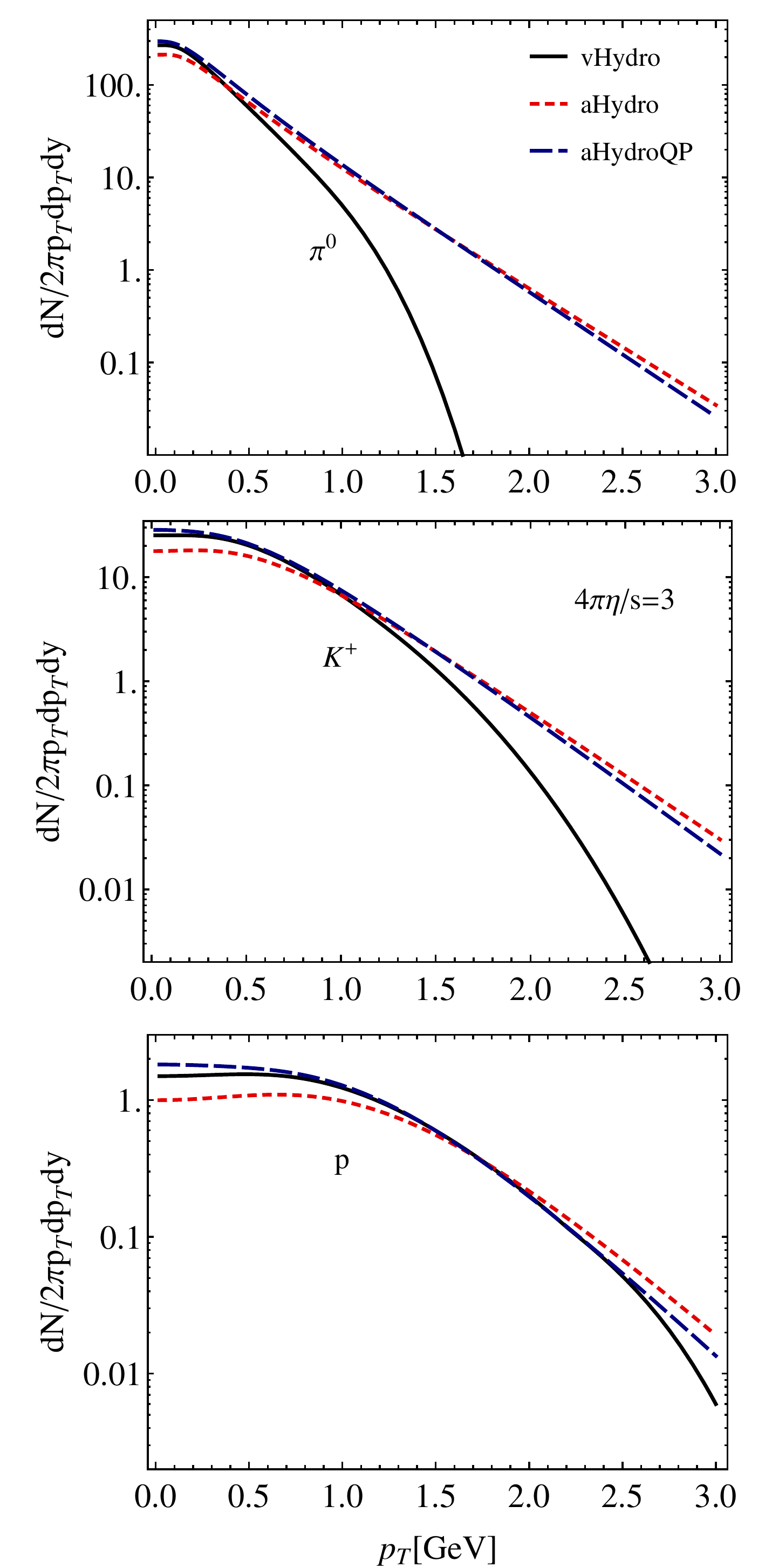}
\caption{Same as Fig.~\ref{fig:spectrum1}, except here the shear viscosity to entropy density ratio was taken to be $ 4 \pi \eta / s =3 $.}
\label{fig:spectrum3}
\end{figure}

\begin{figure}[t]
\includegraphics[width=0.9\linewidth]{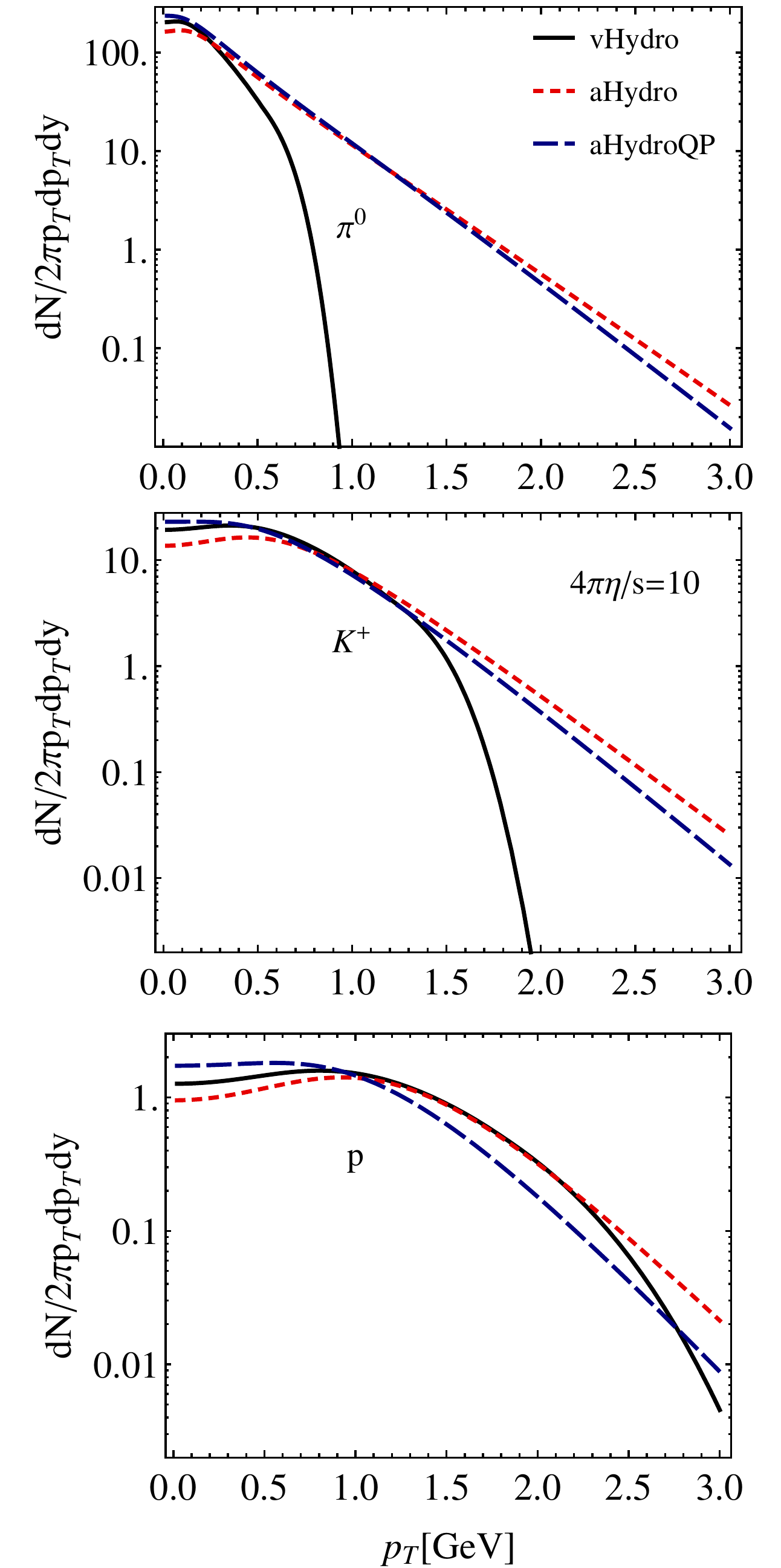}
\caption{Same as Fig.~\ref{fig:spectrum1}, except here the shear viscosity to entropy density ratio was taken to be $ 4 \pi \eta / s =10 $.}
\label{fig:spectrum10}
\end{figure}

\begin{figure}[t]
\vspace{-1mm}
\includegraphics[width=0.89\linewidth]{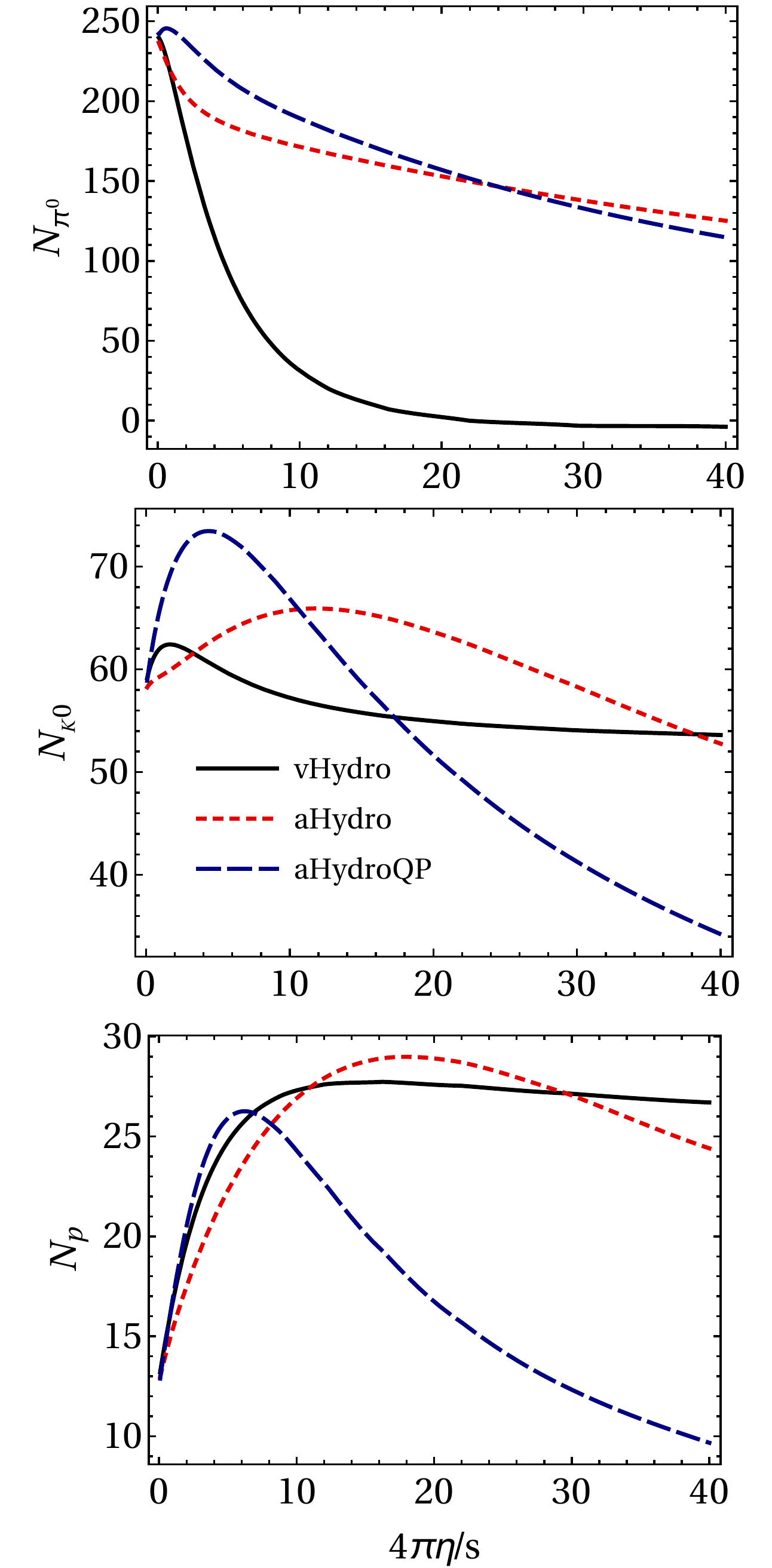}
\caption{Comparison of the total number of pions ($\pi^0$), kaons ($K^0$), and protons as a function of $ 4 \pi \eta / s $ obtained using aHydroQP, aHydro, and  vHydro. }
\label{fig:numberPKPr}
\end{figure}

\begin{figure}[t]
\vspace{1mm}
\includegraphics[width=0.89\linewidth]{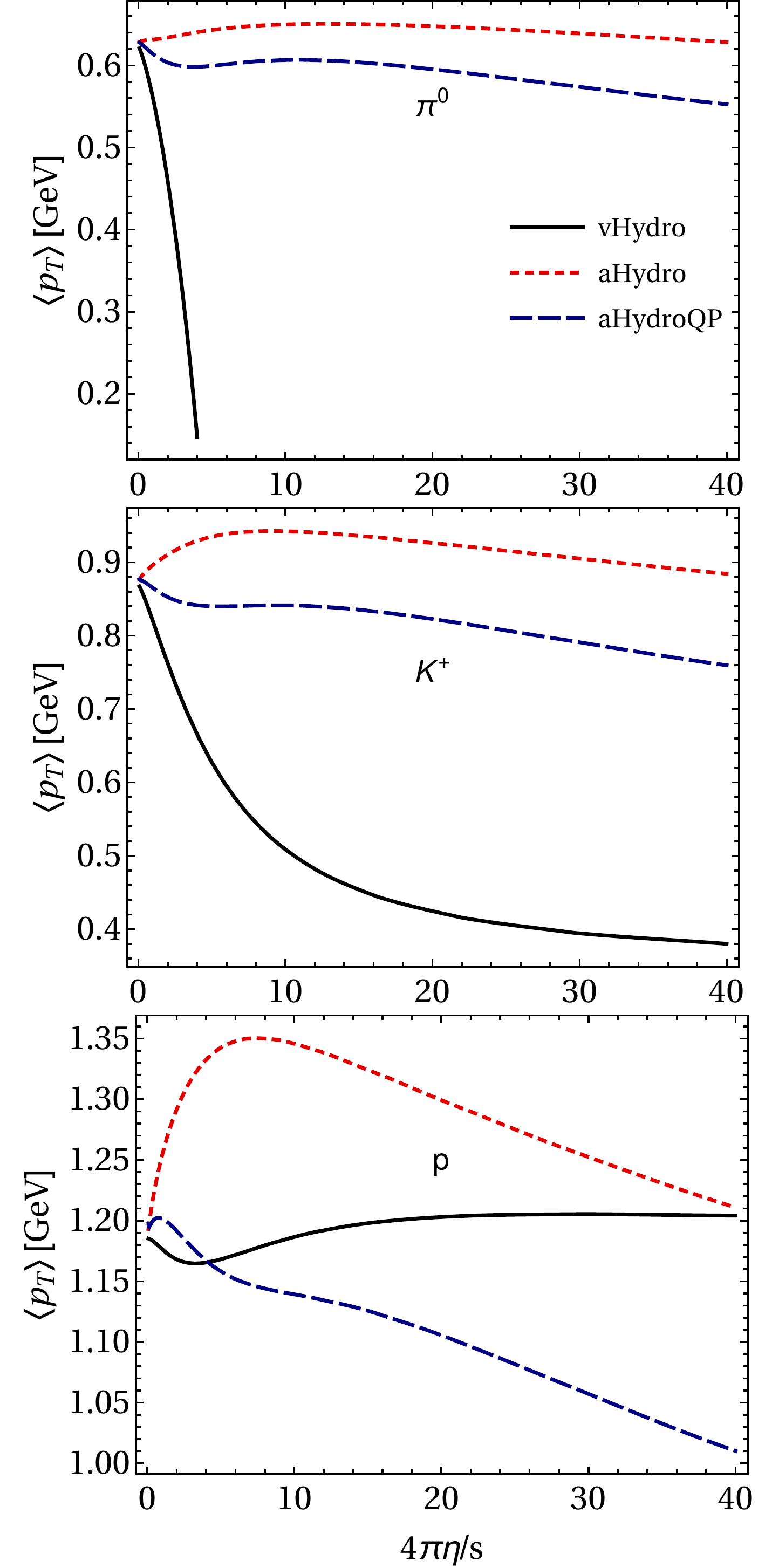}
\caption{Comparison of the neutral pions, kaons ($K^+$), and protons average transverse momentum as a function of $ 4 \pi \eta / s $ obtained using aHydroQP, aHydro, and vHydro.}
\label{fig:ptavg}
\end{figure}

\begin{figure}[t]
\centerline{\includegraphics[width=0.9\linewidth]{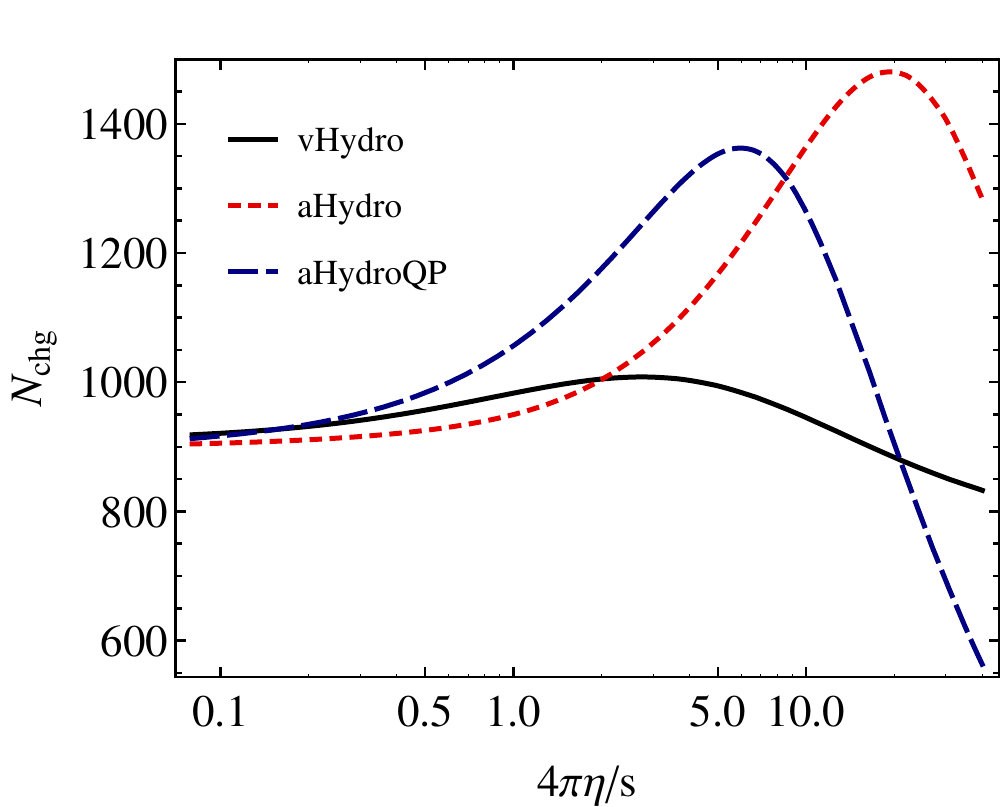}\hspace{5mm}}
\caption{The number of total charged particles as a function of $ 4 \pi \eta / s $ obtained using aHydroQP, aHydro, and  vHydro.}
\label{fig:numberchg}
\end{figure}

\begin{figure}[t]
\includegraphics[width=0.9\linewidth]{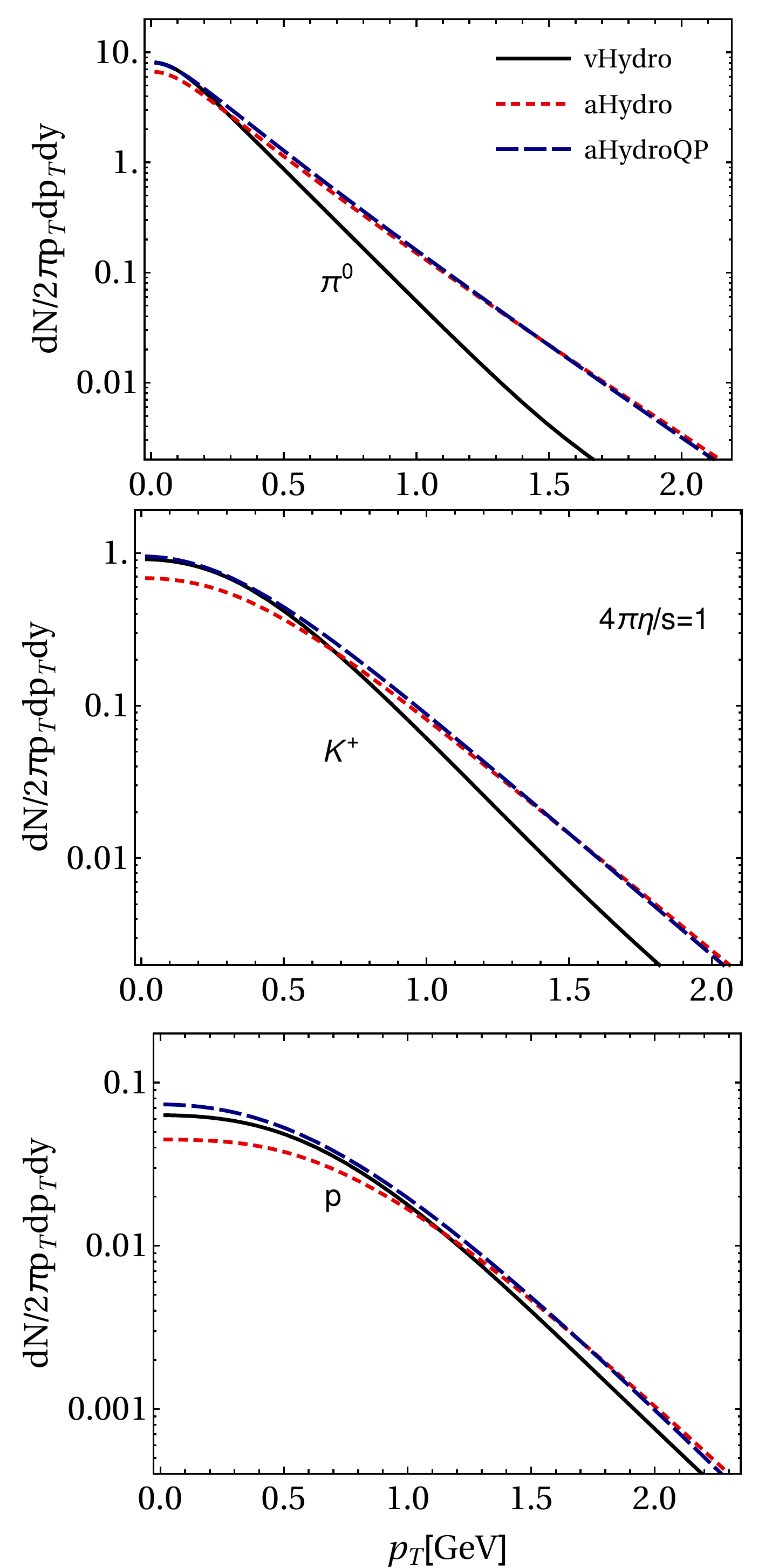}
\caption{Comparisons of the p-Pb neutral pions, kaons ($K^+$), and protons spectra as a function of transverse momentum $ p_T$ obtained using aHydroQP, aHydro, and  vHydro. The shear viscosity to entropy density ratio is $ 4 \pi \eta / s =1$. }
\label{fig:spectrumpPb1}
\end{figure}

\begin{figure}[t]
\includegraphics[width=0.9\linewidth]{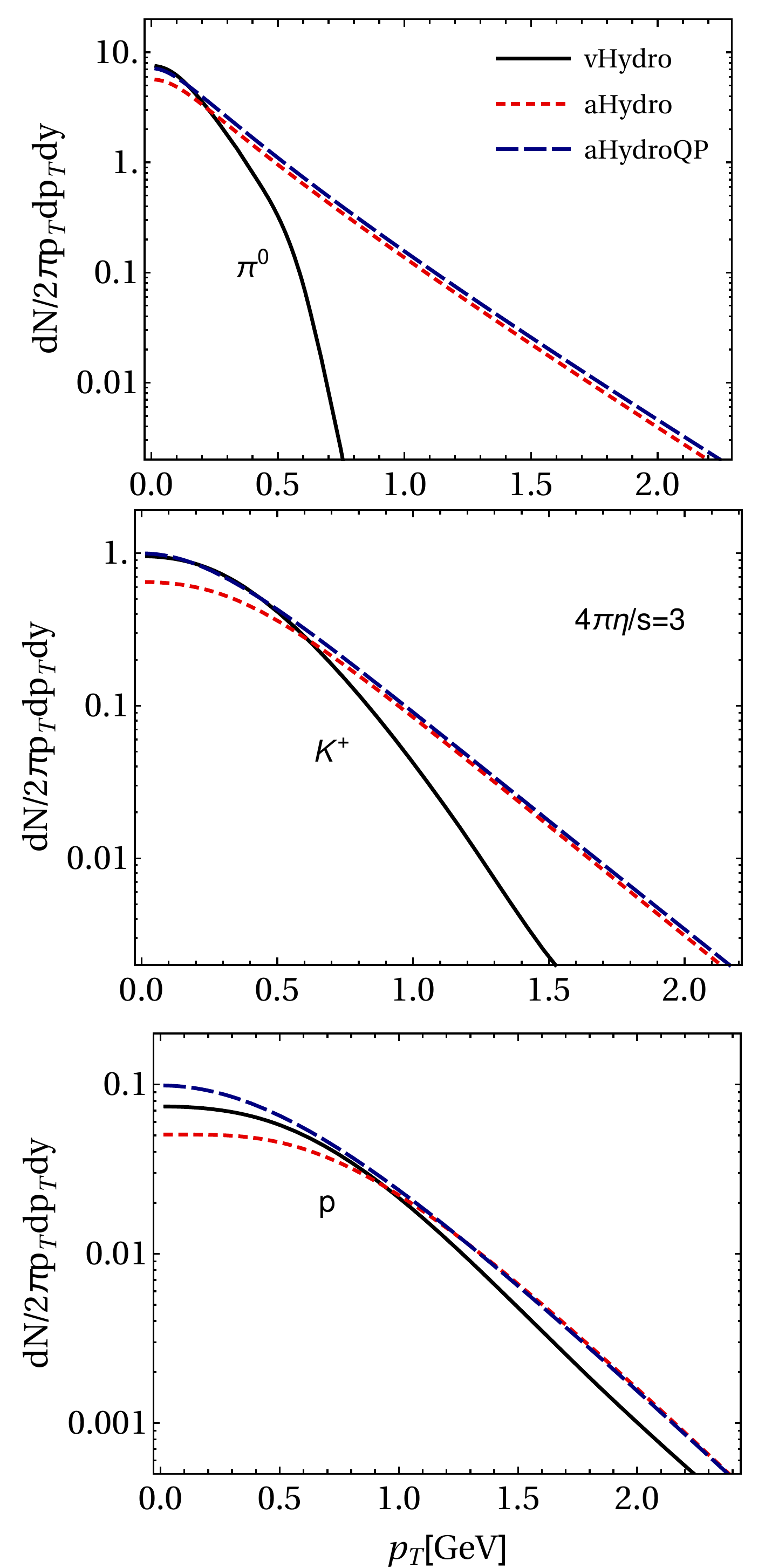}
\caption{Same as Fig.~\ref{fig:spectrumpPb1}, except here the shear viscosity to entropy density ratio was taken to be $ 4 \pi \eta / s =3 $. }
\label{fig:spectrumpPb3}
\end{figure}

\section{Results}
\label{sec:results}

We now turn to our numerical results.  We present comparisons of results obtained using the dynamical equations of anisotropic hydrodynamics presented in Sec.~\ref{sec:dynamical-eqs} and the second-order viscous hydrodynamics equations from Denicol et al. \cite{Denicol:2012cn,Denicol:2014vaa}.  For details about the vHydro equations solved herein we refer the reader to App.~\ref{app:vhydro}.

\paragraph*{Pb-Pb collisions:}

For all results presented in this section we use smooth Glauber wounded-nucleon overlap to set the initial energy density.  As our test case we consider Pb-Pb collisions with $\sqrt{s_{\rm NN}} = $ 2.76 GeV.  The inelastic nucleon-nucleon scattering cross-section is taken to be $\sigma_{\rm NN} = $ 62 mb.  For the aHydro results, we use 200 points in the radial direction with a lattice spacing of $\Delta r = 0.15$ fm and temporal step size of $\Delta \tau =$ 0.01 fm/c.  For the vHydro results, we use 600 points in the radial direction with a lattice spacing of $\Delta r = 0.05$ fm and temporal step size of $\Delta \tau =$ 0.001 fm/c.\footnote{We found that the vHydro code was more sensitive to the spatial lattice spacing and required a smaller temporal step size for stability.}   In all cases, we use fourth-order Runge-Kutta integration for the temporal updates and fourth-order centered differences for the evaluation of all spatial derivatives.\footnote{Since the initial conditions considered herein are smooth, naive centered differences generally suffice.}  We take the central initial temperature to be $T_0 = 600$ MeV at $\tau_0 = 0.25$ fm/c and assume that the system is initially isotropic, i.e. $\alpha_x(\tau_0) =\alpha_y(\tau_0) = \alpha_z(\tau_0) = 1$ for anisotropic hydrodynamics and $\pi^{\mu\nu}(\tau_0)=\Pi(\tau_0)=0$ for second-order viscous hydrodynamics.  We take the freeze-out temperature to be $T_{\rm eff} = T_{\rm FO} = $ 150 MeV in all cases shown.  For the freeze-out we use 371 hadronic resonances ($M_{\rm hadron} \leq 2.6  \; {\rm GeV}$), with the masses, spins, etc. taken from the SHARE table of hadronic resonances \cite{Torrieri:2004zz,Torrieri:2006xi,Petran:2013dva}.  We do not perform resonance feed down, hence all spectrum shown herein are primordial spectra.

In Figs.~\ref{fig:spectrum1}-\ref{fig:spectrum10} we present our results for the primordial pion, kaon, and proton spectra produced for $4\pi\eta/s = 1, 3,$ and 10, respectively.  In each case, we have held the initial conditions fixed and only varied $\eta/s$.  In each of these figures the solid black line is the result from standard second-order viscous hydrodynamics (vHydro), the red short-dashed line is the result from standard anisotropic hydrodynamics (aHydro), and the blue long-dashed line is the result from quasiparticle anisotropic hydrodynamics (aHydroQP).  As can be seen from Fig.~\ref{fig:spectrum1}, for $4\pi\eta/s=1$, both aHydro approaches are in good agreement over the entire $p_T$ range shown with the largest differences occurring at low momentum.  The second-order viscous hydrodynamics result, however, shows a significant downward curvature in the pion spectrum resulting in many fewer high-$p_T$ pions.  The trend is the same for the kaon and proton spectra, however, for the larger mass hadrons the downturn is less severe.  We note that although we plot only up to $p_T = 3$ GeV, these plots can be extended to larger $p_T$, in which case one finds that eventually the primordial pion spectrum predicted by vHydro becomes negative, which is clearly unphysical.  This can be seen more clearly in Figs.~\ref{fig:spectrum3} and \ref{fig:spectrum10} which show the same results for $4\pi\eta/s = 3$ and 10, respectively.  As these figures demonstrate, for larger $\eta/s$ the vHydro primordial particle spectra become unphysical at lower momenta. For example, for $4\pi\eta/s=3$ the differential pion spectrum goes negative at $p_T \sim 1.6$ GeV while for $4\pi\eta/s=10$ it goes negative at $p_T \sim 0.9$ GeV.

This behavior is a result of the bulk-viscous correction to the one-particle distribution function specified in Eq.~(\ref{eq:deltafbulk}).  We have checked that if we neglect the bulk-viscous correction to the distribution function, then the resulting spectra are positive definite in the range of $p_T$ shown in Figs.~\ref{fig:spectrum1}-\ref{fig:spectrum10}.  We have verified that this is a known issue with the bulk-viscous correction in the second-order viscous hydrodynamics approach.  The same form for the bulk correction (\ref{eq:deltafbulk}) was used by the authors of Ref.~\cite{Ryu:2015vwa} and they also observed a downward curvature turning into negatively-valued spectra at large $p_T$ \cite{SchenkeDenicolPrivate}.  This problem does not occur in either aHydro approach because the one-particle distribution function is positive-definite by construction in this framework.

Next, we turn to a discussion of Fig.~\ref{fig:numberPKPr}.  In this figure we plot the total number of $\pi^0$'s (top), $K^0$'s (middle), and $p$'s (bottom) obtained by integrating the differential yields over transverse momentum as a function of $4\pi\eta/s$.  The line styles are the same as in Figs.~\ref{fig:spectrum1}-\ref{fig:spectrum10}.  From this figure we see that at small $\eta/s$ all approaches are in agreement, however, at large $\eta/s$ the three methods can give dramatically different results.  We, in particular, note that the number of pions from vHydro drops much quicker than the two aHydro approaches.  This is primarily due to the fact that in vHydro one sees a negative number of pions at large $p_T$.  As shown in Fig.~\ref{fig:numberPKPr} (top), the total number of pions predicted by vHydro goes negative at $4\pi \eta/s \sim 22$.  This is a signal of the complete breakdown of viscous hydrodynamics and should come as no surprise since this approach is intended to be applicable only in the case of small $\eta/s$ and $p_T$.  Finally, we note that although aHydro and aHydroQP are in reasonably good agreement for the pion and kaon spectra, we see a rather strong dependence on the way the EoS is implemented in the proton spectra and, hence, the total number of primordial protons.

As an additional way to compare the three methods considered, in Fig.~\ref{fig:ptavg} we present the average $p_T$ for $\pi^0$'s (top), $K^+$'s (middle), and $p$'s (bottom).  The line styles are the same as in Figs.~\ref{fig:spectrum1}-\ref{fig:spectrum10}.  From this figure we see that both aHydro approaches predict a weak dependence of the pion and kaon $\langle p_T \rangle$ on the assumed value of $\eta/s$, whereas vHydro predicts a much more steep decrease in $\langle p_T \rangle$ for the pions and kaons.  Once again, we see that the vHydro $\langle p_T \rangle$ for pions becomes negative for $4\pi\eta/s \gtrsim 5$.  This rapid decrease stems directly from the negativity of the pion spectra at high $p_T$ which is more important in this case since the integrand is more sensitive to the high-$p_T$ part of the spectra.  Finally, we note that once again the proton spectra and hence $\langle p_T \rangle$ for protons is sensitive to the way in which the EoS is implemented when comparing the two aHydro approaches.

Finally, in Fig.~\ref{fig:numberchg} we plot the total number of charged particles as a function of $4\pi\eta/s$ predicted by each of the three approaches considered.  Once again the line styles are the same as in Figs.~\ref{fig:spectrum1}-\ref{fig:spectrum10}.  As this figure demonstrates, for small $\eta/s$ all three frameworks are in agreement and approach the ideal result as $\eta/s$ tends to zero.  All three frameworks predict that $N_{\rm chg}$ at first increases, then reaches a maximum, and then begins to decrease.  The precise turnover point depends on the method with the lowest turnover seen using vHydro around $4 \pi \eta/s \sim 3$; however, this turnover is due in large part to the fact that the total pion number drops precipitously in vHydro, eventually becoming negative, as we pointed out in the discussion of Fig.~\ref{fig:numberPKPr}.  As a result, one cannot trust the large $\eta/s$ predictions of vHydro.

\paragraph*{p-Pb collisions:}
We now turn to the case of an asymmetric collision between a proton and a nucleus.  For this purpose, we use the same parameters as used for the Pb-Pb collisions considered previously, except for p-Pb we use a lower initial temperature of $T_0 = 400$ MeV . In addition, for the aHydro results, we use a lattice spacing of $\Delta r = 0.06$ fm and for vHydro results, we use a lattice spacing of $\Delta r = 0.02$ fm.  The smaller lattice spacings simply reflect the smaller system size of the QGP created in a p-Pb collision.

In Figs.~\ref{fig:spectrumpPb1}-\ref{fig:spectrumpPb3} we present our results for the primordial pion, kaon, and proton spectra produced for $4\pi\eta/s = 1,$ and 3, respectively in p-Pb collisions. As we can see from Fig.~\ref{fig:spectrumpPb1}, both aHydro approaches are in a good agreement at high $p_T$, however, there are some quantitative differences at low $p_T$. Comparing the low $p_T$ difference with that seen in Pb-Pb collisions, we find that there is a larger variation in the p-Pb spectra comparing aHydro and aHydroQP.  This variation is a bit worrisome since it indicates a kind of theoretical uncertainty in the aHydro approach.  Importantly, however, we mention that the two aHydro results are quite different than the vHydro result.  For p-Pb collisions, the vHydro result shows the same behavior seen in the Pb-Pb collisions, namely that the particle spectra goes negative at high $p_T$.   In Fig.~\ref{fig:spectrumpPb3}, one can clearly see the unphysical behavior in the vHydro primordial spectra.  As a result, there is even larger theoretical uncertainty associated with applications of vHydro to p-Pb collisions.

\section{Conclusions and Outlook}
\label{sec:conclusions}
In this paper, we compared three different viscous hydrodynamics approaches:  aHydro, aHydroQP, and vHydro.  For all three cases we included both shear and bulk-viscous effects using the relaxation time approximation scattering kernel.  In the standard aHydro approach one uses the standard method for imposing an equation of state in anisotropic hydrodynamics, which is to obtain conformal equations and then break the conformality only when introducing the EoS itself.  In aHydroQP, one takes into account the breaking of conformality at the outset by modeling the QGP as a quasiparticle gas with a single temperature-dependent mass $m$ which is fit to available lattice data for the EoS.  Finally, for our comparisons with viscous hydrodynamics we used the formalism of Denicol et al \cite{Denicol:2012cn,Denicol:2014vaa} specialized to the case of relaxation time approximation.

For each method, we specialized to the case of 1+1d boost-invariant and azimuthally-symmetric collisions using smooth Glauber initial conditions.  We specialized to this case because of the computational intensity of the aHydroQP approach which requires real-time evaluate of complicated multi-dimensional integrals which are functions of all three anisotropy parameters and the local temperature-dependent mass.  Using the resulting numerical evolution, we then extracted fixed energy density freeze-out hypersurfaces in each case and implemented the scheme-appropriate freeze-out to hadrons allowing us to have an apples-to-apples comparison between the three different approaches.  We found that the primordial particle spectra, total number of charged particles, and average transverse momentum predicted by the three methods agree well for small shear viscosity to entropy density ratio, $\eta/s$, but differ at large $\eta/s$.  Our most important finding was that when using standard viscous hydrodynamics, the bulk-viscous correction can drive the primordial particle spectra negative at large $p_T$.  Such a behavior is not seen in either anisotropic hydrodynamics approach, irrespective of the value of $\eta/s$.

Looking to the future, it is feasible to extend the aHydroQP approach to (3+1)d, however, this will be numerically intensive and require parallelization to implement fully.  One possibility is to use polynomial fits to parametrize the various massive ${\cal H}$-functions necessary instead of evaluating them on-the-fly in the code.  If this is possible, then aHydroQP could become a viable alternative to the standard method of implementing the EoS in aHydro and, since it takes into account the non-conformality of the system from the beginning, this could give us an idea of the theoretical uncertainty associated with the EoS method used in phenomenological applications.  For now, this work will serve as a reference point for possible differences between the two approaches to imposing the aHydro EoS.

\acknowledgments{We thank G. Denicol and B. Schenke for useful discussions. M. Strickland and M. Nopoush were supported by the U.S. Department of Energy, Office of Science, Office of Nuclear Physics under Awards No.~DE-SC0013470.  M. Alqahtani was supported by a PhD fellowship from the University of Dammam.}
\begin{widetext}
\appendix 

\section{Basis Vectors}
\label{app:basis}
One can define the general basis vectors in the lab frame (LF) by performing the Lorentz transformation necessary to go from local rest frame to the LF.  The transformation required can be constructed using a longitudinal boost $\vartheta$ along the beam axis, followed by a rotation $\varphi$ around the beam axis, and finally a transverse boost by $\theta_\perp$ along the $x$-axis~\cite{Florkowski:2011jg,Martinez:2012tu}. This parametrization gives
\ba
u^\mu &\equiv& \left(\cosh\theta_\perp \cosh\vartheta,\sinh\theta_\perp\cos\varphi,\sinh\theta_\perp\sin\varphi,\cosh\theta_\perp \sinh\vartheta\right) ,  \nonumber \\
X^\mu &\equiv& \left(\sinh\theta_\perp \cosh\vartheta,\cosh\theta_\perp\cos\varphi,\cosh\theta_\perp\sin\varphi,\sinh\theta_\perp \sinh\vartheta\right) , \nonumber  
\ea
\ba  
Y^\mu &\equiv& \left(0,-\sin\varphi,\cos\varphi,0\right) , \nonumber  \\
Z^\mu &\equiv& \left(\sinh\vartheta,0,0,\cosh\vartheta\right) ,
\label{eq:basisvecs}
\ea
where the three fields $\vartheta$, $\varphi$, and $\theta_\perp$ are functions of Cartesian Milne coordinates $(\tau,x,y,\varsigma)$. Introducing another parametrization by using the temporal and transverse components of flow velocity, one has
\begin{equation}
\begin{split}
u_0&=\cosh\theta_\perp\, , \\
u_x&= u_\perp \cos\varphi\, , \\
u_y&= u_\perp \sin\varphi\, ,
\end{split}
\quad\Longrightarrow\qquad
\begin{split}
u^\mu &\equiv (u_0 \cosh\vartheta,u_x,u_y,u_0 \sinh\vartheta) \, , \\
X^\mu &\equiv \Big(u_\perp\cosh\vartheta,\frac{u_0 u_x}{u_\perp},\frac{u_0 u_y}{u_\perp},u_\perp\sinh\vartheta\Big) ,  \\ 
Y^\mu &\equiv \Big(0,-\frac{u_y}{u_\perp},\frac{u_x}{u_\perp},0\Big)  ,  \\
Z^\mu &\equiv (\sinh\vartheta,0,0,\cosh\vartheta ) \, ,
\label{eq:4vectors}
\end{split}
\end{equation}
where $u_\perp\equiv \sqrt{u_x^2+u_y^2}=\sqrt{u_0^2-1} = \sinh\theta_\perp$.
For a boost-invariant and azimuthally-symmetric system, one can simplify the basis vectors by identifying $\vartheta=\varsigma$ and $\varphi=\phi$ where $\varsigma$ and $\phi$ are the spacetime rapidity and the azimuthal angle, respectively.  In this case, the basis vectors (\ref{eq:basisvecs}) simplify to
\ba
u^\mu &=& (\cosh\theta_\perp \cosh\varsigma,\sinh\theta_\perp \cos\phi,\sinh\theta_\perp \sin\phi,\cosh\theta_\perp \sinh\varsigma) \, , \nonumber \\
X^\mu &=& (\sinh\theta_\perp \cosh\varsigma, \cosh\theta_\perp \cos\phi,\cosh\theta_\perp \sin\phi,\sinh\theta_\perp \sinh\varsigma) \, , \nonumber \\
Y^\mu &=&(0,-\sin\phi,\cos\phi,0)\, , \nonumber \\
Z^\mu &=&(\sinh\varsigma,0,0,\cosh\varsigma)\, .
\label{eq:basis-1+1}
\ea
%
\section{Second-order viscous hydrodynamics}
\label{app:vhydro}
Similar to anisotropic hydrodynamics, the viscous hydrodynamics dynamical equations can be obtained by taking moments of the Boltzmann equation. The energy-momentum tensor in the case when the bulk correction is not ignored is
\be
T^{\mu\nu}={\cal E} u^\mu u^\nu - \Delta^{\mu\nu}({\cal P}+\Pi) + \pi^{\mu\nu} \, .
\label{eq:T-expan-vHydro}
\ee
 Taking the first and second moments of Boltzmann equation one obtains \cite{Denicol:2014vaa,Denicol:2012cn}
\ba
({\cal E}+{\cal P}+\Pi)D_u u^\mu&=&\nabla^\mu ({\cal P}+\Pi)-\Delta^\mu_{\nu} \nabla_\sigma
\pi^{\nu \sigma}+\pi^{\mu \nu} D_u u_\nu\, ,
\label{eq:vhyd1-1}\\
D_u {\cal E} &=& - ({\cal E}+{\cal P}+\Pi) \theta_u+ \pi^{\mu \nu}
\sigma_{\mu\nu}\, ,
\label{eq:vhyd2-1}\\
\tau _{\Pi }D_{u}{\Pi}+\Pi &=&-\zeta \theta_u -\delta _{\Pi \Pi }\Pi \theta_u
+\varphi _{\mathrm{1}}\Pi ^{2}+\lambda _{\Pi \pi }\pi ^{\mu \nu }\sigma
_{\mu \nu }+\varphi _{\mathrm{3}}\pi ^{\mu \nu }\pi _{\mu \nu } \, ,
\label{eq:vhyd3-1} \\
\tau _{\pi }\Delta^{\mu\nu}_{\alpha\beta} D_u \pi^{\alpha \beta}+\pi ^{\mu \nu } &=&2\eta
\sigma ^{\mu \nu }+2\tau _{\pi }\pi _{\alpha }^{\langle \mu }\omega ^{\nu
\rangle \alpha }-\delta _{\pi \pi }\pi ^{\mu \nu }\theta_u +\varphi _{\mathrm{7%
}}\pi _{\alpha }^{\langle \mu }\pi ^{\nu \rangle \alpha }-\tau _{\pi \pi
}\pi _{\alpha }^{\langle \mu }\sigma ^{\nu \rangle \alpha } 
+\,\lambda _{\pi \Pi }\Pi \sigma ^{\mu \nu }+\varphi _{6}\Pi \pi ^{\mu \nu
}\,, \hspace{5mm}
\label{eq:vhyd4-1}
\ea
where ${\cal E}\equiv{\cal E}_{\rm eq}$ and ${\cal P}\equiv {\cal P}_{\rm eq}$ are the equilibrium (isotropic) energy density and pressure, respectively, $\tau_\pi$ and $\tau_{\Pi}$  are the shear and bulk relaxation time, respectively, and $\tau_{\pi\pi}$ is the shear-shear-coupling transport coefficient.  The various notations used are 
\begin{equation}
\begin{split}
d_\mu u^\nu &\equiv \partial_\mu u^\nu+\Gamma_{\mu\alpha}^{\nu} u^\alpha \, ,  \\[.4ex]
D_u &\equiv u_\mu d^\mu \, ,  \\[1.5ex]
\theta_u &\equiv \nabla _{\mu }u^{\mu } \, , \\[1.5ex]
\nabla^\mu &\equiv \Delta^{\mu \nu} d_\nu \, ,
\end{split}
\quad\quad
\begin{split}
\sigma^{\mu\nu}&\equiv \nabla^{\langle\mu} u^{\nu\rangle}\, ,  \\[.2ex]
A^{\langle \mu
\nu \rangle }&\equiv  \Delta _{\alpha \beta }^{\mu \nu }A^{\alpha \beta }\, ,  \\[-.3ex]
\Delta _{\alpha \beta }^{\mu \nu } &\equiv \frac{1}{2}\Big(\Delta _{\alpha }^{\mu
}\Delta _{\beta }^{\nu }+\Delta _{\beta }^{\mu }\Delta _{\alpha }^{\nu
}-\frac{2}{3}\Delta ^{\mu \nu }\Delta _{\alpha \beta }\Big) \, ,  \\
\omega^{\mu \nu} &\equiv \frac{1}{2} ( \nabla^\mu u^\nu - \nabla^\nu u^\mu) \, .
\label{eq:identities1}
\end{split}
\end{equation}
The non-vanishing Christoffel symbols for polar Milne coordinates are $\Gamma^\tau_{\varsigma\varsigma}=\tau$, $\Gamma^\varsigma_{\varsigma\tau}=1/\tau$, $\Gamma^r_{\phi\phi}=-r$, and $\Gamma^\phi_{r\phi}=1/r$.
Also, we note that for the smooth initial conditions considered herein in 1+1d, the vorticity
tensor vanishes. As shown in Ref. \cite{Molnar:2013lta}, the terms $%
\varphi _{\mathrm{1}}\Pi ^{2}$, $\varphi _{\mathrm{3}}\pi ^{\mu \nu }\pi
_{\mu \nu }$, $\varphi _{6}\Pi \pi ^{\mu \nu }$, and $\varphi _{\mathrm{7}%
}\pi _{\alpha }^{\langle \mu }\pi ^{\nu \rangle \alpha }$ appear only
because the collision term is nonlinear in the single-particle distribution
function. In the case of the RTA, the collision term is assumed to be linear
in the single-particle distribution function and one has $\varphi _{\mathrm{1}}=\varphi _{\mathrm{3}}=\varphi _{%
\mathrm{6}}=\varphi _{\mathrm{7}}=0${.} In this case,
the shear and bulk relaxation times, $\tau _{\pi }$ and $\tau _{\Pi }$,
respectively, are equal to the microscopic relaxation time $\tau _{\mathrm{eq%
}}$, i.e., $\tau _{\Pi }=\tau _{\pi }=\tau _{\mathrm{eq}}$   
\cite{Denicol:2014vaa}. 
The coefficients appearing in the equation for the bulk and shear corrections are~\cite{Denicol:2014vaa}
\begin{equation}
\begin{split}
\frac{\zeta }{\tau _{\Pi }} &=\bigg( \frac{1}{3}-c_{s}^{2}\bigg) (\mathcal{%
E}+\mathcal{P})-\frac{2}{9}(\mathcal{E}-3\mathcal{P})\,, \\
\frac{\delta _{\Pi \Pi }}{\tau _{\Pi }} &=1-c_{s}^{2}\,, \\
\frac{\lambda _{\Pi \pi }}{\tau _{\Pi }} &=\frac{1}{3}-c_{s}^{2}\,,
\end{split}
\quad\quad
\begin{split}
\frac{\eta }{\tau _{\pi }} &=\frac{4}{5}\mathcal{P}+\frac{1}{15}(\mathcal{E}%
-3\mathcal{P})\,, \\
\frac{\delta _{\pi \pi }}{\tau _{\pi }} &=\frac{4}{3}\,, \\
\frac{\tau _{\pi \pi }}{\tau _{\pi }} &=\frac{10}{7}\,, \\
\frac{\lambda _{\pi \Pi }}{\tau _{\pi }} &=\frac{6}{5}\,,
\end{split}
\end{equation}
with $c_s^2\equiv d{\cal P}/d{\cal E}$ being the speed of sound squared and $ \tau_{\rm eq} =15 \bar{\eta}(\mathcal{E}+\mathcal{P})/(\mathcal{E}+9\mathcal{P})/T$.  
\subsection*{1+1d viscous hydrodynamics equations of motion}
In the boost-invariant and azimuthally-symmetric case, one has $u^\mu = (u^\tau,u^r,0,0)$ and, as a result, $v \equiv \tanh\theta_\perp = u^r/u^\tau$.  In addition, for this case, the shear tensor has the following form
\be
\pi^{\mu\nu}=
\begin{pmatrix}
\pi^{\tau\tau}&\pi^{\tau r}&0&0\\ \pi^{\tau r}&\pi^{rr}&0&0\\0&0&\pi^{\phi\phi}&0\\0&0&0&\pi^{\varsigma\varsigma}
\end{pmatrix} .
\label{eq:pi-mat-1}
\ee
In this case, expanding Eqs.~(\ref{eq:vhyd1-1}), (\ref{eq:vhyd2-1}), and (\ref{eq:vhyd3-1}) in polar Milne coordinates one obtains six equations where only five of them are independent 
\ba
({\cal E}+{\cal P}+\Pi)D_u u^\tau &=&-(u^r)^2\Big[\partial_\tau( {\cal P}+\Pi)
-d_\nu \pi^\nu_\tau\Big] 
-u^\tau u^r \Big[\partial_r( {\cal P}+\Pi)-d_\nu \pi^\nu_r\Big] , \label{eq:ut}\\
({\cal E}+{\cal P}+\Pi)D_u u^r &=& -u^\tau u^r\Big[\partial_\tau ({\cal P}+\Pi)-d_\nu \pi^\nu_\tau\Big]
-(u^\tau)^2 \Big[\partial_r ( {\cal P}+\Pi)-d_\nu\pi^\nu_r\Big] , \label{eq:ur} \\
D_u{\cal E} &=&\!-({\cal E}+{\cal P}+\Pi) \theta_u 
\!-\pi^r_r (1-v^2)^2\nabla^{\langle r} u^{r\rangle}\!\!-r^2\, \pi^\phi_\phi \nabla^{\langle\phi} u^{\phi\rangle}
\!\!-\tau^2\,\pi^\varsigma_\varsigma \nabla^{\langle\varsigma} u^{\varsigma \rangle}\!, \hspace{.8cm} \label{eq:e}
\ea
and
\ba
\tau_{\Pi}D_u \Pi+\Pi &=& -\zeta \theta_u - \delta_{\Pi \Pi} \Pi \theta_u -\lambda_{\Pi \pi}\Big[2 r^2 \pi^\phi_\phi \nabla^{\langle\phi}
u^{\phi \rangle} + 2 \tau^2 \pi^\varsigma_\varsigma \nabla^{\langle\varsigma}
u^{\varsigma \rangle} 
\nonumber \\ 
&+& r^2 \pi^\varsigma_\varsigma \nabla^{\langle\phi}
u^{\phi \rangle} + \tau^2 \pi^\phi_\phi \nabla^{\langle\varsigma}
u^{\varsigma \rangle}    \Big]  \, , \hspace{0.8cm} \label{eq:Pi-ff}  \\
\tau_{\pi} D_u \pi^\phi_\phi + \pi^\phi_\phi &=& -2 r^2 \eta \nabla^{\langle\phi}u^{\phi \rangle} - \delta_{\pi \pi} \pi^\phi_\phi \theta_u - \frac{\tau_{\pi \pi}}{3}\Big[- r^2 \pi^\phi_\phi \nabla^{\langle\phi}
u^{\phi \rangle} + 2 \tau^2 \pi^\varsigma_\varsigma \nabla^{\langle\varsigma}
u^{\varsigma \rangle} 
\nonumber \\ 
&+& r^2 \pi^\varsigma_\varsigma \nabla^{\langle\phi}
u^{\phi \rangle}  +  \tau^2 \pi^\phi_\phi \nabla^{\langle\varsigma}
u^{\varsigma \rangle} \Big]-r^2 \lambda_{\pi \Pi} \Pi \nabla^{\langle\phi}
u^{\phi \rangle}  \, ,  \label{eq:pi-ff}  \\
\tau_{\pi} D_u \pi^\varsigma_\varsigma + \pi^\varsigma_\varsigma &=& -2 \tau^2 \eta \nabla^{\langle\varsigma}u^{\varsigma \rangle} - \delta_{\pi \pi} \pi^\varsigma_\varsigma \theta_u - \frac{\tau_{\pi \pi}}{3}\Big[2 r^2 \pi^\phi_\phi \nabla^{\langle\phi}
u^{\phi \rangle} -  \tau^2 \pi^\varsigma_\varsigma \nabla^{\langle\varsigma}
u^{\varsigma \rangle} 
\nonumber \\ 
&+& r^2 \pi^\varsigma_\varsigma \nabla^{\langle\phi}
u^{\phi \rangle}  +  \tau^2 \pi^\phi_\phi \nabla^{\langle\varsigma}
u^{\varsigma \rangle} \Big]-\tau^2 \lambda_{\pi \Pi} \Pi \nabla^{\langle\varsigma}
u^{\varsigma \rangle}  \, ,\label{eq:pi-vv}
\ea
where 
\ba
- d_\nu \pi^\nu_\tau &=& v^2 \partial_\tau \pi^r_r+v \partial_r \pi^r_r+\pi^r_r\Big[\partial_\tau v^2+\partial_r v+\frac{v^2}{\tau}
+\frac{v}{r}\Big]+\frac{1}{\tau}\pi^\varsigma_\varsigma \, , \label{eq:dpi-tau}\\
d_\nu \pi^\nu_r &=& v\, \partial_\tau \pi^r_r+\partial_r \pi^r_r+\pi^r_r \Big[\partial_\tau v+\frac{v}{\tau}+\frac{2-v^2}{r}\Big]
+\frac{1}{r} \pi^\varsigma_\varsigma \, . \label{eq:dpi-r}
\ea
In addition, one needs the following identities
\ba 
\nabla^{\langle r} u^{r \rangle} &=& - \partial_r u^r- u^r D_u u^r +\frac{1}{3} (u^\tau)^2 \theta_u \, ,\label{eq:nablar-ur}
\\ 
r^2 \nabla^{\langle \phi} u^{\phi \rangle} &=&- \frac{u^r}{r}+\frac{1}{3}\theta_u \, , \label{eq:nablavph-uph} 
\ea
\ba
\tau^2 \nabla^{\langle \varsigma} u^{\varsigma \rangle} &=&- \frac{u^\tau}{\tau}+\frac{1}{3}\theta_u \, , \label{eq:nablav-uv} \\
\theta_u\equiv\nabla_\alpha u^\alpha &=&d_\alpha u^\alpha = \partial_\tau u^\tau+\partial_r u^r+\frac{u^\tau}{\tau}+\frac{u^r}{r}\, ,
\ea
where $\pi^r_\tau=-v\, \pi^r_r$ and $\pi^\phi_\phi=-\pi^\varsigma_\varsigma-(1-v^2) \pi^r_r$ which are a consequence of the transversality of the shear-stress tensor, $u_\mu \pi^{\mu\nu}=0$.  This system of equations has to be closed by providing an equation of state (EoS), e.g. ${\cal P}_{\rm eq}={\cal P}_{\rm eq}({\cal E}_{\rm eq})$.  For the numerical results presented in the body of the manuscript, we use the lattice-based EoS specified in Sec.~\ref{subsec:eos}.
\subsection*{Viscous hydrodynamics freeze-out}
The distribution function on the freeze-out hypersurface can be computed assuming that there is a linear correction to the equilibrium one due to shear and bulk viscosities \cite{Rose:2014fba,Teaney:2003kp}
\ba
f(p,x)&=& f_{\rm eq}(p,x) + \delta f_{\rm shear}(p,x)+ \delta f_{\rm bulk}(p,x)\, ,
\label{eq:vfreeze}
\ea
where
\ba
\delta f_{\rm shear}(p,x)&=&f_{\rm eq} (1 - a f_{\rm eq}) \frac{p_\mu p_\nu \pi^{\mu\nu}}{2({\cal E}+{\cal P})T^2} \, , 
\label{eq:deltashear} \\
\delta f_{\rm bulk}(p,x)&=& -f_{\rm eq}(1 - a f_{\rm eq}) \bigg[ \frac{m_i^2}{3\,  p_\mu u^\mu } - \Big(\frac{1}{3}-c_s^2\Big)p_\mu u^\mu \bigg]\frac{\Pi}{C_\Pi} \, , 
\label{eq:deltafbulk}
\ea
with
\be
C_\Pi = \frac{1}{3} \sum_{i=1}^N m_i^2 (2s_i+1) \int \frac{d^3 p}{(2 \pi)^3 E_i}  f_{\rm eq}(1 - a f_{\rm eq})\bigg[ \frac{m_i^2}{3\,  p_\mu u^\mu } - \Big(\frac{1}{3}-c_s^2\Big)p_\mu u^\mu \bigg]  \, ,
\ee
where $m_i$ is the hadron mass, $s_i$ is the hadron spin, and $N$ is the number of hadrons included in the freezeout.
Using tensor transformations applied to Eq.~(\ref{eq:ptl-mom}) the components of the four-momentum in polar Milne coordinates are
\ba
p_\tau &=&p_t\cosh\varsigma-p_z\sinh\varsigma=m_\perp\cosh(y-\varsigma) \, , \nonumber \\
p_r&=&p_x \cos\phi+p_y\sin\phi=p_\perp\cos(\phi-\varphi) \, , \nonumber \\
p_\phi&=&-p_x\frac{\sin\phi}{r}+p_y\frac{\cos\phi}{r}=-\frac{p_\perp}{r}\sin(\phi-\varphi)\, , \nonumber \\
p_\varsigma&=&-p_t\frac{\sinh\varsigma}{\tau}+p_z\frac{\cosh\varsigma}{\tau}=\frac{m_\perp}{\tau}\sinh(y-\varsigma) \, .
\label{eq:p-con}
\ea
Using Eq.~(\ref{eq:pi-mat-1}) and expanding $p_\mu p_\nu\pi^{\mu\nu}$ in polar Milne coordinates one has
\ba 
p_\mu p_\nu \pi^{\mu\nu}=&-&\bigg(\frac{\pi^\phi_\phi+\pi^\varsigma_\varsigma}{v^2-1}\bigg)\Big(m_\perp v \cosh(y-\varsigma)-p_\perp\cos(\phi-\varphi)\Big)^2 \nonumber \\
&-&\pi^\phi_\phi p^2_\perp\sin^2(\phi-\varphi)-\pi^\varsigma_\varsigma m^2_\perp\sinh^2(y-\varsigma) \, .
\ea
\section{Explicit formulas for derivatives}
\label{app:identities}
In this section, we introduce the notations used in hydrodynamics dynamical equations. In the case of boost-invariant and azimuthally-symmetric flow one can use (\ref{eq:basis-1+1}) to obtain 
\begin{equation}
\setlength{\jot}{1pt}
\begin{split}
D_u&=u^\mu\partial_\mu=\cosh\theta_\perp \partial_\tau+\sinh\theta_\perp\partial_r\,,\\[1.3ex]
D_x&=X^\mu\partial_\mu=\sinh\theta_\perp\partial_\tau+\cosh\theta_\perp\partial_r\,,\\[.1ex]
D_y&=Y^\mu\partial_\mu=\frac{1}{r}\partial_\phi\,, \\
D_z&=Z^\mu\partial_\mu=\frac{1}{\tau}\partial_\varsigma\,,
\end{split}
\quad\quad
\begin{split}
\theta_u&=\partial_\mu u^\mu=\cosh\theta_\perp\Big(\frac{1}{\tau}+\partial_r\theta_\perp\Big)+\sinh\theta_\perp\Big(\frac{1}{r}+\partial_\tau\theta_\perp\Big) ,\\
\theta_x&=\partial_\mu X^\mu=\sinh\theta_\perp\Big(\frac{1}{\tau}+\partial_r\theta_\perp\Big)+\cosh\theta_\perp\Big(\frac{1}{r}+\partial_\tau\theta_\perp\Big) ,\\[.1ex]
\theta_y&= \partial_\mu Y^\mu=0\, ,\\[1.7ex]
\theta_z&= \partial_\mu Z^\mu=0\,.
\end{split}
\end{equation}

\section{special functions}
\label{app:h-functions}
In this section, we provide definitions of the special functions appearing in the body of the text. We start by introducing 
\ba
 {\cal H}_2(y,z) 
&=& \frac{y}{\sqrt{y^2-1}} \left[ (z^2+1)
\tanh^{-1} \sqrt{\frac{y^2-1}{y^2+z^2}} + \sqrt{(y^2-1)(y^2+z^2)} \, \right] ,
\label{eq:H2} \\
{\cal H}_{2T}(y,z) 
&=& \frac{y}{(y^2-1)^{3/2}}
\left[\left(z^2+2y^2-1\right) 
\tanh^{-1}\sqrt{\frac{y^2-1}{y^2+z^2}}
-\sqrt{(y^2-1)(y^2+z^2)} \right] , \hspace{1cm}
\label{eq:H2T} \\
{\cal H}_{2L}(y,z) &=& \frac{y^3}{(y^2-1)^{3/2}}
\left[
\sqrt{(y^2-1)(y^2+z^2)}-(z^2+1)
\tanh^{-1}\sqrt{\frac{y^2-1}{y^2+z^2}} \,\,\right], 
\label{eq:H2L}
\ea
and
\ba
{\cal H}_{2x1}(y,z)&=&\frac{1}{(y^2-1)}\Big[\frac{2 (y^2+z^2){\cal H}_{2L}(y,z)}{(1+z^2)}-y^2{\cal H}_{2T}(y,z)\Big] , \\
{\cal H}_{2x2}(y,z)&=&\frac{y^2}{(y^2-1)}\Big[2 {\cal H}_{2L}(y,z)-{\cal H}_{2T}(y,z)\Big]\, , \\
{\cal H}_{2B}(y,z)&\equiv & {\cal H}_{2T}(y,z)+ \frac{{\cal H}_{2L}(y,z)}{y^2}=\frac{2}{\sqrt{y^2-1}}\tanh^{-1} \sqrt{\frac{y^2-1}{y^2+z^2}} \, .
\ea
Derivatives of these functions satisfy the following relations
\begin{equation}
\begin{split}
\frac{\partial {\cal H}_2(y,z)}{\partial y}&=\frac{1}{y}\Big[{\cal H}_2(y,z)+{\cal H}_{2L}(y,z)\Big] , \\
\frac{\partial {\cal H}_2(y,z)}{\partial z}&=\frac{1}{z}\Big[{\cal H}_2(y,z)-{\cal H}_{2L}(y,z)-{\cal H}_{2T}(y,z)\Big] , 
\end{split}
\quad\quad
\begin{split}
\frac{\partial {\cal H}_{2T}(y,z)}{\partial y}&=\frac{1}{y(y^2-1)}\Big[2{\cal H}_{2L}(y,z)-{\cal H}_{2T}(y,z)\Big] , \\
\frac{\partial {\cal H}_{2T}(y,z)}{\partial z}&=\frac{-2z}{y^2(1+z^2)}{\cal H}_{2L}(y,z) .
\end{split}
\end{equation}
%
\subsection{Massive Case} 
\label{subapp:h-functions-1}
The ${\cal H}$-functions appearing in definitions of components of the energy-momentum tensor in the massive case are 
\ba
{\cal H}_3({\boldsymbol\alpha},\hat{m}) &\equiv &  \tilde{N} \alpha_x \alpha_y
\int_0^{2\pi} d\phi \, \alpha_\perp^2 \int_0^\infty d\hat{p} \, \hat{p}^3  f_{\rm eq}\!\left(\!\sqrt{\hat{p}^2 + \hat{m}^2}\right) {\cal H}_2\!\left(\frac{\alpha_z}{\alpha_\perp},\frac{\hat{m}}{\alpha_\perp \hat{p}} \right),
\label{eq:h3gen}
\\
{\cal H}_{3x}({\boldsymbol\alpha},\hat{m}) &\equiv &  \tilde{N}\alpha_x^3 \alpha_y
\int_0^{2\pi} d\phi \, \cos^2\phi \int_0^\infty d\hat{p} \, \hat{p}^3  f_{\rm eq}\!\left(\!\sqrt{\hat{p}^2 + \hat{m}^2}\right) {\cal H}_{2T}\!\left(\frac{\alpha_z}{\alpha_\perp},\frac{\hat{m}}{\alpha_\perp \hat{p}} \right),
\;\;\;\;
\label{eq:h3xgen}
\\
{\cal H}_{3y}({\boldsymbol\alpha},\hat{m}) &\equiv &  \tilde{N}\alpha_x \alpha_y^3
\int_0^{2\pi} d\phi \, \sin^2\phi \int_0^\infty d\hat{p} \, \hat{p}^3  f_{\rm eq}\!\left(\!\sqrt{\hat{p}^2 + \hat{m}^2}\right) {\cal H}_{2T}\!\left(\frac{\alpha_z}{\alpha_\perp},\frac{\hat{m}}{\alpha_\perp \hat{p}} \right) ,
\label{eq:h3ygen}
\\
{\cal H}_{3T}({\boldsymbol\alpha},\hat{m}) &\equiv &  \frac{1}{2} \Big[ {\cal H}_{3x}({\boldsymbol\alpha},\hat{m}) + {\cal H}_{3y}({\boldsymbol\alpha},\hat{m}) \Big] ,
\label{eq:h3tgen}
\\
{\cal H}_{3L}({\boldsymbol\alpha},\hat{m}) &\equiv &  \tilde{N} \alpha_x \alpha_y
\int_0^{2\pi} d\phi \, \alpha_\perp^2 \int_0^\infty d\hat{p} \, \hat{p}^3  f_{\rm eq}\!\left(\!\sqrt{\hat{p}^2 + \hat{m}^2}\right) {\cal H}_{2L}\!\left(\frac{\alpha_z}{\alpha_\perp},\frac{\hat{m}}{\alpha_\perp \hat{p}} \right) ,
\label{eq:h3lgen} 
\\
{\cal H}_{3B}({\boldsymbol\alpha},\hat{m}) &\equiv &  \tilde{N}\alpha_x\alpha_y
\int_0^{2\pi} d\phi\int_0^\infty d\hat{p} \, \hat{p} f_{\rm eq}\!\left(\!\sqrt{\hat{p}^2 + \hat{m}^2}\right) {\cal H}_{\rm 2B}\!\left(\frac{\alpha_z}{\alpha_\perp},\frac{\hat{m}}{\alpha_\perp \hat{p}} \right) ,
\label{eq:h3Bgen} 
\ea
%
The relevant derivatives are
\begin{equation}
\begin{split}
\frac{\partial {\cal H}_3}{\partial\alpha_x}&=\frac{1}{\alpha_x}({\cal H}_3+{\cal H}_{3x})\,, \\
\frac{\partial {\cal H}_3}{\partial\alpha_y}&=\frac{1}{\alpha_y}({\cal H}_3+{\cal H}_{3y})\,, \\
\frac{\partial {\cal H}_3}{\partial\alpha_z}&=\frac{1}{\alpha_z}({\cal H}_3+{\cal H}_{3L})\,, \\
\frac{\partial {\cal H}_3}{\partial\hat{m}}&=\frac{1}{\hat{m}}({\cal H}_3-{\cal H}_{3L}-2{\cal H}_{3T}-{\cal H}_{3m})\,, \\
\end{split}
\quad\quad
\begin{split}
\frac{\partial {\cal H}_{3x}}{\partial\alpha_x}&=\frac{1}{\alpha_x}({3\cal H}_{3x}-{\cal H}_{3x1})\,, \\
\frac{\partial {\cal H}_{3x}}{\partial\alpha_y}&=\frac{1}{\alpha_y}({\cal H}_{3x}-{\cal H}_{3x2})\,, \\
\frac{\partial {\cal H}_{3x}}{\partial\alpha_z}&=\frac{1}{\alpha_z}{\cal H}_{3x3}\,, \\
\frac{\partial {\cal H}_{3x}}{\partial\hat{m}}&=\frac{1}{\hat{m}}( {\cal H}_{3m1}-{\cal H}_{3m2})\,,
\end{split}
\end{equation}
with
\ba
{\cal H}_{3x1}({\boldsymbol\alpha},\hat{m}) &\equiv &  \tilde{N}\frac{\alpha_x^5 \alpha_y}{\alpha_z^2}
\int_0^{2\pi}\!\!d\phi \, \cos^4\phi \int_0^\infty d\hat{p} \, \hat{p}^3  f_{\rm eq}\!\left(\!\sqrt{\hat{p}^2 + \hat{m}^2}\right) {\cal H}_{2x1}\!\left(\frac{\alpha_z}{\alpha_\perp},\frac{\hat{m}}{\alpha_\perp \hat{p}} \right),\label{eq:h3x1gen}
\\
{\cal H}_{3x2}({\boldsymbol\alpha},\hat{m}) &\equiv &  \tilde{N}\frac{\alpha_x^3 \alpha_y^3}{\alpha_z^2}
\int_0^{2\pi} \!\!d\phi \,\cos^2\phi \sin^2\phi  \int_0^\infty d\hat{p} \, \hat{p}^3  f_{\rm eq}\!\left(\!\sqrt{\hat{p}^2 + \hat{m}^2}\right) {\cal H}_{2x1}\!\left(\frac{\alpha_z}{\alpha_\perp},\frac{\hat{m}}{\alpha_\perp \hat{p}} \right), \,\,\,\,\,\,\,
\;\;\;\;\label{eq:h3x2gen}
\\
{\cal H}_{3x3}({\boldsymbol\alpha},\hat{m}) &\equiv &  \tilde{N} \frac{\alpha_x^3 \alpha_y}{\alpha_z^2}
\int_0^{2\pi} d\phi \, \alpha_\perp^2 \, \cos^2\phi \int_0^\infty d\hat{p} \, \hat{p}^3  f_{\rm eq}\!\left(\!\sqrt{\hat{p}^2 + \hat{m}^2}\right) {\cal H}_{2x2}\!\left(\frac{\alpha_z}{\alpha_\perp},\frac{\hat{m}}{\alpha_\perp \hat{p}} \right),
\;\;\;\;\label{eq:h3x3gen}
\\
{\cal H}_{3m}({\boldsymbol\alpha},\hat{m}) &\equiv &  \tilde{N}\alpha_x\alpha_y\hat{m}^2
\int_0^{2\pi} d\phi \, \alpha_\perp^2\int_0^\infty d\hat{p} \,\hat{p}^3 \frac{f_{\rm eq}\!\left(\!\sqrt{\hat{p}^2 + \hat{m}^2}\right)}{\sqrt{\hat{p}^2+\hat{m}^2}}   {\cal H}_2\!\left(\frac{\alpha_z}{\alpha_\perp},\frac{\hat{m}}{\alpha_\perp \hat{p}} \right) ,
\label{eq:h3mgen} 
\\
{\cal H}_{3m1}({\boldsymbol\alpha},\hat{m}) &\equiv & {\cal H}_{3x1}({\boldsymbol\alpha},\hat{m})+{\cal H}_{3x2}({\boldsymbol\alpha},\hat{m})-{\cal H}_{3x3}({\boldsymbol\alpha},\hat{m}) \, ,
\;\;\;\;\label{eq:h3m1gen}
\\
{\cal H}_{3m2}({\boldsymbol\alpha},\hat{m}) &\equiv &   \tilde{N}\alpha_x^3 \alpha_y \hat{m}^2
\int_0^{2\pi} d\phi \, \cos^2\phi \int_0^\infty d\hat{p} \, \hat{p}^3  \frac{f_{\rm eq}\!\left(\!\sqrt{\hat{p}^2 + \hat{m}^2}\right)}{\sqrt{\hat{p}^2 + \hat{m}^2}}  {\cal H}_{2T}\!\left(\frac{\alpha_z}{\alpha_\perp},\frac{\hat{m}}{\alpha_\perp \hat{p}} \right) ,
\;\;\;\;\label{eq:h3m2gen}
\ea
where $\alpha_\perp^2 \equiv \alpha_x^2 \cos^2\phi  + \alpha_y^2 \sin^2\phi$.
\subsection{Massless Case}
\label{subapp:h-functions-2}
The $\hat{\cal H}$-functions used in definition of bulk variables in the standard (massless) case are
\ba
\hat{{\cal H}}_3 ({\boldsymbol\alpha})&\equiv & \frac{1}{24 \pi \tilde{N}} \lim_{m\rightarrow 0}{\cal H}_3({\boldsymbol\alpha},\hat{m}) =  \frac{1}{4 \pi} \alpha_x \alpha_y
\int_0^{2\pi} d\phi \, \alpha_\perp^2 \bar{{\cal H}}_2\Big(\frac{\alpha_z}{\alpha_\perp} \Big)\, ,
\label{eq:h30}
\\
\hat{{\cal H}}_{3x} ({\boldsymbol\alpha})&\equiv & \frac{1}{8 \pi \tilde{N}} \lim_{m\rightarrow 0}{\cal H}_{3x}({\boldsymbol\alpha},\hat{m}) =   \frac{3}{4 \pi}\alpha_x^3 \alpha_y
\int_0^{2\pi} d\phi \, \cos^2\phi \, \bar{{\cal H}}_{2T}\Big(\frac{\alpha_z}{\alpha_\perp} \Big)\, ,
\label{eq:h3x0}
\\
\hat{{\cal H}}_{3y} ({\boldsymbol\alpha})&\equiv & \frac{1}{8 \pi \tilde{N}} \lim_{m\rightarrow 0}{\cal H}_{3y}({\boldsymbol\alpha},\hat{m}) =   \frac{3}{4 \pi}\alpha_x \alpha_y^3
\int_0^{2\pi} d\phi \, \sin^2\phi \, \bar{{\cal H}}_{2T}\Big(\frac{\alpha_z}{\alpha_\perp} \Big)\, ,
\label{eq:h3y0}
\\
\hat{{\cal H}}_{3L} ({\boldsymbol\alpha})&\equiv & \frac{1}{8 \pi \tilde{N}} \lim_{m\rightarrow 0}{\cal H}_{3L}({\boldsymbol\alpha},\hat{m}) =   \frac{3}{4 \pi} \alpha_x \alpha_y
\int_0^{2\pi} d\phi \, \alpha_\perp^2 \bar{{\cal H}}_{2L}\Big(\frac{\alpha_z}{\alpha_\perp} \Big)\,.
\label{eq:h3l0}
\ea
\end{widetext}
\bibliography{1+1Quasiparticle}

\end{document}